\documentclass[pre,amsmath,amssymb,groupedaddress]{revtex4}

\usepackage{graphicx,subfigure}
\usepackage{bm}
\usepackage{mathtools}
\usepackage{verbatim}
\usepackage{dsfont}
\usepackage{amsmath}
\usepackage{amssymb}
\usepackage[T1]{fontenc}
\usepackage{ae,aecompl}

\newcommand{\xh}{{\hat x}}
\newcommand{\yh}{{\hat y}}
\newcommand{\zh}{{\hat z}}

\newcommand{\av}{{\bf a}}
\newcommand{\cv}{{\bf c}}

\newcommand{\grad}{{\bm{\nabla}}}

\newcommand{\jv}{{\bf j}}

\newcommand{\rv}{{\bf r}}
\newcommand{\rvv}{{\vec r}}
\newcommand{\hv}{{\vec h}}
\newcommand{\Av}{{\bf A}}

\newcommand{\xv}{{\bf x}}
\newcommand{\kv}{{\bf k}}

\newcommand{\uv}{{\bf u}}

\newcommand{\qv}{{\bf q}}

\newcommand{\nh}{{\hat n}}

\newcommand{\eh}{{\hat{\bf e}}}

\def\mm#1{\underline{\underline{{#1}}}}

\newcommand{\oh}{{\frac{1}{2}}}

\newcommand{\tRe}{{\text{Re}}}
\newcommand{\cH}{{\mathcal H}}

\newcommand{\be}{\begin{equation}}
\newcommand{\ee}{\end{equation}}
\newcommand{\bea}{\begin{eqnarray}}
\newcommand{\eea}{\end{eqnarray}}
\newcommand{\bse}{\begin{subequations}}
\newcommand{\ese}{\end{subequations}}
\def\rf#1{(\ref{#1})}
\def\rfs#1{Eq.~\rf{#1}}

\usepackage{graphicx} 

\begin{document}
\title
{Critical Matter}
\author{Leo Radzihovsky}
\affiliation{Department of Physics, University of Colorado, Boulder,
  CO 80309, USA}

\date{\today}

\begin{abstract}
  I review a class of novel ordered states of ``critical matter'',
  that exhibit strongly fluctuating universal power-law orders,
  controlled by an infra-red attractive, non-Gaussian fixed point. I
  will illustrate how RG methods pioneered by Wilson and Fisher can be
  used to deduce critical phenomenology of such critical phases,
  resembling that of a critical point of second order phase
  transitions, but requiring no fine tuning.
\end{abstract}

\maketitle


\section{Introduction}
\label{intro}

Michael Fisher is a towering figure in theoretical physics, most
notably recognized for his seminal development (in collaboration with
Ken Wilson and building on the works of Kadanoff, Migdal, Widom,
Larkin, Pokrovsky, and others) of the renormalization group
(RG)\cite{epsilonWilsonFisher, MEFrmp74, WilsonKogutPR} and its
numerous early applications to critical phenomena of continuous phase
transitions\cite{ChaikinLubensky}.  In these works Michael transformed
Wilson's deep ideas\cite{MEFrmp74, WilsonKogutPR} into a practical and
powerful theoretical tool for controlled
calculations\cite{epsilonWilsonFisher} (complementing more formal
developments in quantum field theory\cite{ZinnJustin}) and
demonstrated its power through numerous seminal applications to
critical phenomena.  With this, he elevated RG into a central
calculation tool in modern theoretical physics.
\begin{figure}[htbp]
\hspace{0in}\includegraphics*[width=0.7\textwidth]
{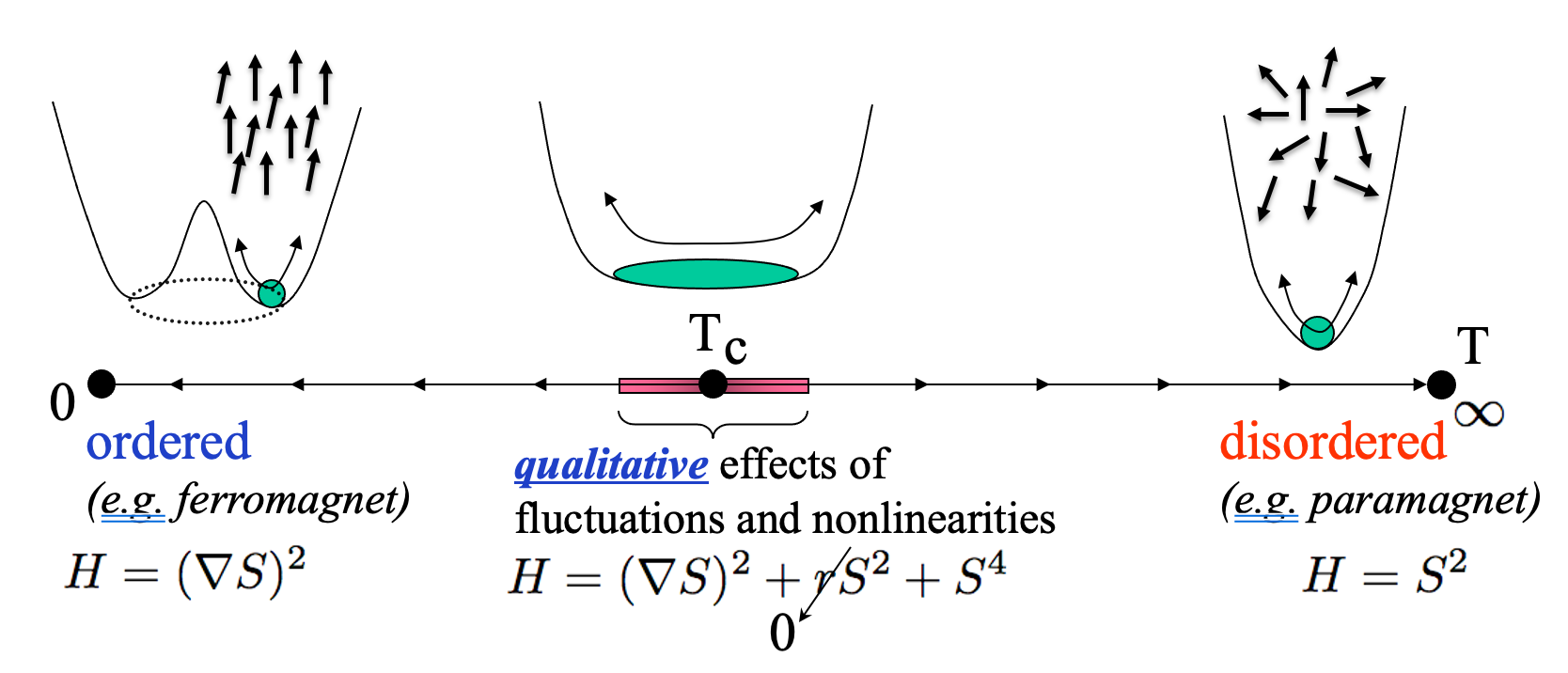}
\caption{Illustration of unimportance of fluctuations inside phases of
  conventional systems, where qualitative effects of thermal
  fluctuations are confined to a vicinity of a critical point.}
\label{conventionalPhaseFixedPointFig}
\end{figure}

Beginning with the earliest seminal analysis of the $\phi^4$ field
theory to describe criticality of the Ising paramagnet-ferromagnet
transition, most applications of RG are used to treat enhanced
fluctuations near a critical point of a continuous phase transition,
where direct perturbation theory breaks down. In such applications a
critical behavior emerges after tuning a set of parameters - e.g.,
temperature, pressure, magnetic field, etc, to a (multi-) critical
point in a phase diagram, as illustrated in
Fig.\ref{conventionalPhaseFixedPointFig}.  It is only then that
fluctuations and nonlinearities (interactions) become important (below
the upper-critical dimension) and universal asymptotic critical
behavior emerges, e.g., power-law, scale-free correlation functions
and thermodynamic responses with universal
exponents\cite{criticalTcomment}.

In contrast to critical points, as illustrated in
Fig.\ref{conventionalPhaseFixedPointFig} phases of matter, by their
very definition are stable to weak generic perturbations that do not
explicitly break their symmetry. They are thus characterized by
infrared {\em attractive} fixed points, and thereby require no
fine-tuning, in contrast to a critical point that separates them.
Fluctuations in ordered Landau phases of matter\cite{Landau} (that is
our focus here) are characterized by their Goldstone modes associated
with the spontaneous breaking of a {\em continuous} symmetry -- Landau
phases that break {\em discrete} symmetries have no interesting
fluctuations. Despite being gapless, generically within ordered phases
Goldstone-mode fluctuations are small, with finite root-mean-squared
fluctuations, latter typically taken as the defining property of the
ordered phase.  In the RG parlance, this corresponds to typical phases
that are controlled by a Gaussian attractive fixed point, as
illustrated in Fig.\ref{conventionalPhaseFixedPointFig}.  Thus,
conventional Landau phases' description is effectively trivial, with
fluctuations (above the lower-critical dimension, where the ordered
phase is stable) only leading to small corrections to their mean-field
description.

\begin{figure}[htbp]
\hspace{0in}\includegraphics*[width=0.7\textwidth]{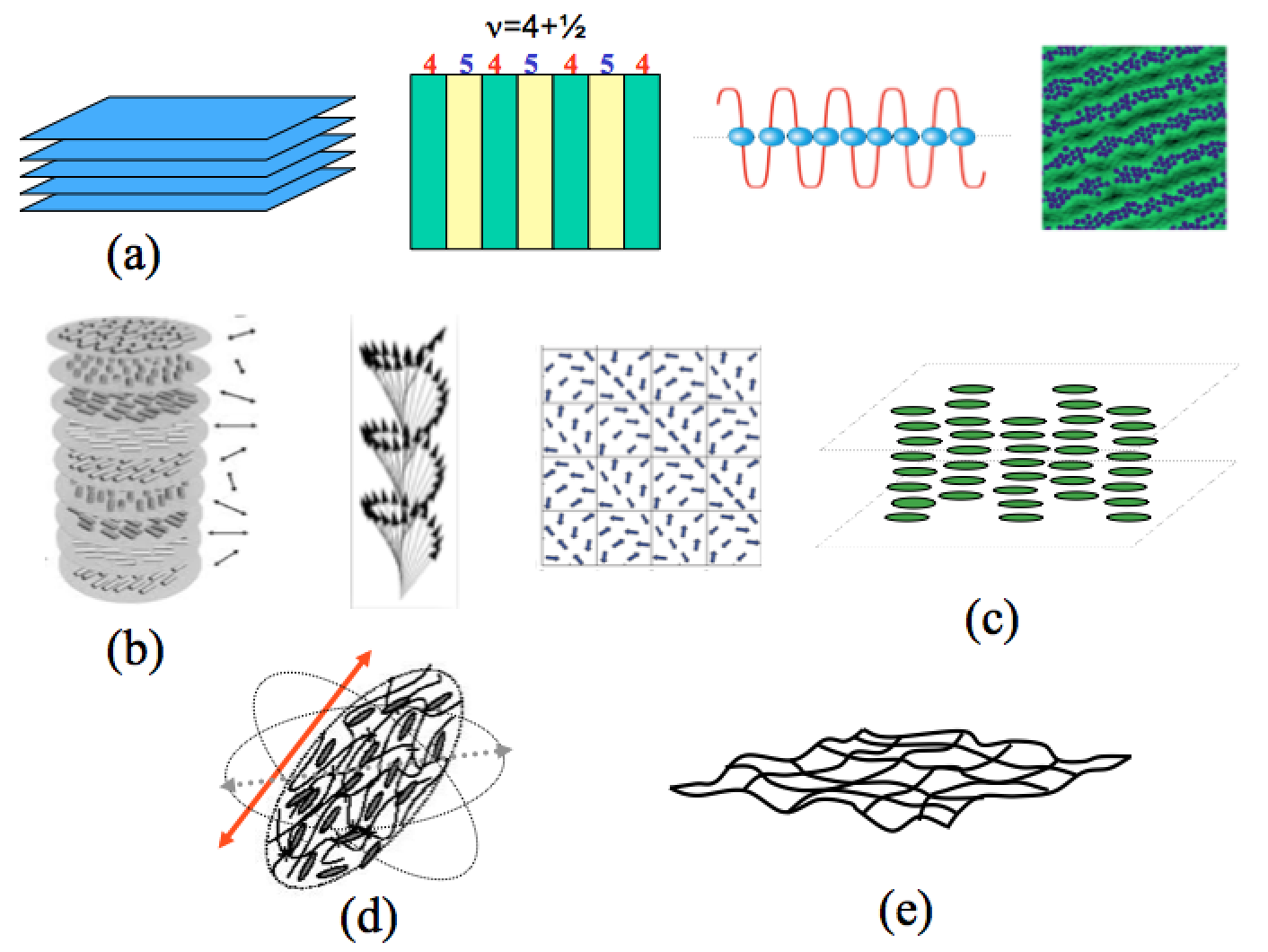}
\caption{Illustration of critical phases: (a) smectics realized as
  conventional liquid crystals\cite{deGennes,ChaikinLubensky,GP,
    RTaerogelPRL, RTaerogelPRB, BRTCaerogelScience}, 2d
  colloids\cite{Glaser}, quantum-Hall
  systems\cite{EisensteinSm, Du99, Fogler,Moessner,MacDonaldFisherSm,FradkinKivelsonPRB,
    LRDorseyQHN} and as striped ``pair-density wave'' FFLO\cite{FF,LO}
  superconducting phases in degenerate atomic gases\cite{ffloLR,
    SRreview,PDWaop}, p-wave resonant\cite{LR_ChoiPRL}, spin-orbit
  coupled\cite{HuiZhai} and frustrated
  superfluids\cite{helicalBosonsHMR22},(b) cholesteric liquid
  crystals\cite{deGennes,ChaikinLubensky, cholestericLubensky,
    cholestericLR} and helical state of frustrated
  magnets\cite{cholestericLR, MnSi,BalentsNature, AParamakanti}, (c)
  columnar liquid crystals and spontaneous vortex
  lattices\cite{PhysicsToday, experiments,
    Varma,SRTcolumnar,magneticSCprl,magneticSCprb}, (d) nematic
  elastomers\cite{bookWarner,XingRadzAOP}, and (e) polymerized
  membranes\cite{JWSmembranes,NelsonPeliti, AronovitzLubensky,DG,
    LeDoussalRadzihovskyPRL, LRphdThesis}.}
\label{criticalPhasesFig}
\end{figure}

\begin{figure}[htbp]
\hspace{0in}\includegraphics*[width=0.7\textwidth]
{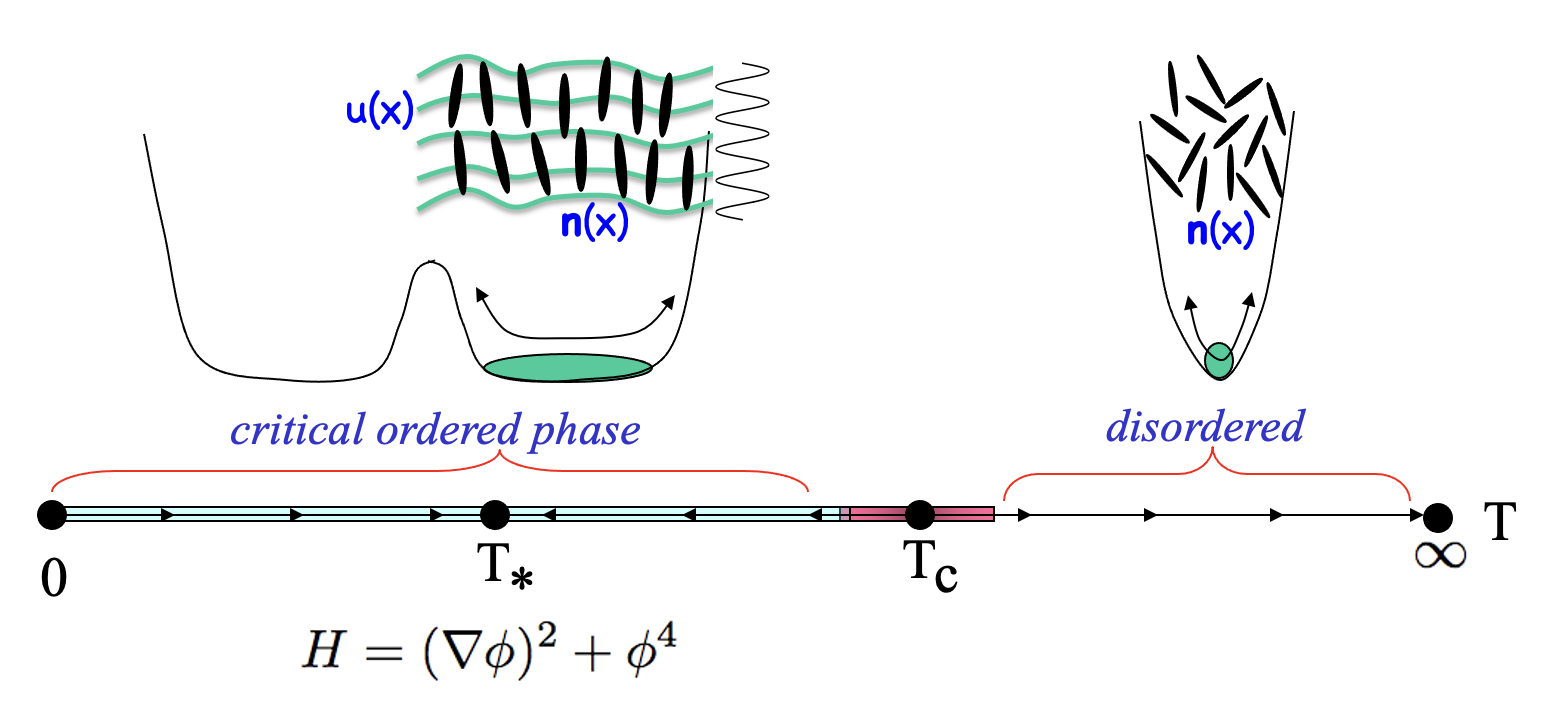}
\caption{Illustration of the importance of fluctuations inside ordered
  phases of ``critical matter'', in which even deep in the ordered
  phase the quartic potential is missing the stabilizing quadratic
  contribution, resulting in divergent fluctuations akin to a critical
  point. The bottom of the figure is a schematic of a renormalization
  group flow, e.g., for a smectic state in $d<3$ dimensions,
  illustrating that at low $T$ it is a ``critical phase'' displaying
  universal power-law phenomenology, controlled by a nontrivial
  infrared-stable fixed point.}
\label{criticalPhaseFixedPointFig}
\end{figure}

In this chapter I instead focus on an exotic class of ordered phases
of matter -- ``critical phases'', e.g., smectics, cholesterics,
columnar phases, membranes, elastomers,\ldots, illustrated in
Fig.\ref{criticalPhasesFig} -- where in stark contrast to conventional
phases, Goldstone-mode fluctuations are divergingly strong,
interacting, and thus are controlled by a {\em non}-Gaussian
infrared attractive fixed point, as illustrated in
Fig.\ref{criticalPhaseFixedPointFig}.
As such the resulting ordered critical phase, while breaking a
continuous symmetry is highly nontrivial and strongly interacting,
with little resemblance of its mean-field cartoon. It is characterized
by universal critical exponents, but in striking constrast to a
critical point requires no fine-tuning, protected by underlying
spontaneously broken symmetries, with importance of strong
fluctuations and nonlinearities extending throughout the ordered
phase.  A generic ingredient of such critical phases (often associated
with spontaneous breaking of {\em continuous} spatial symmetries) is
that underlying symmetry enforces an exact {\em vanishing} of a subset
of elastic moduli, resulting in particularly ``soft'' harmonic
elasticity of the associated Goldstone modes, described by
$m$-Lifshitz models\cite{mLifshitz,BakSOC}.

In subsequent sections of this chapter, I will illustrate above
phenomenology through a set of examples of critical phases, some
illustrated in Fig.\ref{criticalPhasesFig}.  I will begin with a
detailed discussion of a quintessential critical phases -- uniaxially
periodically modulated states that spontaneously break translational
and rotational symmetry, and are realized in numerous physical
contexts, including smectic liquid
crystals\cite{ChaikinLubensky,deGennes, GP},
cholesterics\cite{cholestericLR}, helical magnets and bosons on
frustrated lattices\cite{cholestericLR, MnSi,
  BalentsNature,helicalBosonsHMR22}, nonzero momentum superfluids such
as Fulde-Ferrell-Larkin-Ovchinnikov (FFLO)
superconductors\cite{ffloLR,SRreview}, p-wave resonantly paired
bosons\cite{LR_ChoiPRL}, spin-orbit coupled bosons\cite{HuiZhai}, and
quantum-Hall striped states\cite{EisensteinSm, Du99,
  Fogler,Moessner,MacDonaldFisherSm,FradkinKivelsonPRB}.
In Sec. 3, I will describe a columnar states, that appear in discotic
liquid crystals\cite{ChaikinLubensky,deGennes, SRTcolumnar} and in its
line-crystal analogs, such as putative spontaneous vortex lattice in
magnetic superconductors\cite{magneticSCprl,magneticSCprb}. I will
analyze another rich example of a critical phase -- a thermally
fluctuating tensionless polymerized membrane in
Sec. 4.\cite{JWSmembranes, NelsonPeliti, AronovitzLubensky,DG,
  LeDoussalRadzihovskyPRL} In Sec. 5, I will discuss a nematic
elastomer -- a liquid crystal rubber, that, mathematically is a
three-dimensional amalgam of a smectic and columnar phases, displaying
critical phenomenology throughout its phase\cite{bookWarner,
  XingRadzAOP}.

\section{Uniaxially periodic critical states}
\label{uniaxialSec}
There is a number of states of matter that {\em spontaneously} develop
a periodic modulation along an arbitrarily chosen {\em single}
axis. Strictly speaking this can only take place in a system with
underlying rotational invariance, though can be exhibited over an
extended intermediate regime when explicitly broken rotational
symmetry couples only weakly, as e.g., in low density electronic
systems.\cite{EisensteinSm, Du99}

Beyond conventional liquid crystals, some of the novel systems include
helical states of frustrated and spin-orbit coupled magnets and
bosons\cite{MnSi, BalentsNature,AParamakanti, HuiZhai,
  helicalBosonsHMR22}, and strongly correlated electronic and bosonic
systems (quite surprising for isotropic and point constituents), such
as FFLO\cite{ffloLR} ``striped'' (``pair density wave'')
superconductors\cite{PDWaop,FradkinKivelsonPRB}, finite momentum
superfluids\cite{LR_ChoiPRL, HuiZhai,helicalBosonsHMR22}, and a
two-dimensional electron gas in the quantum Hall regime of half-filled
Landau level \cite{EisensteinSm, Du99, Fogler,Moessner,MacDonaldFisherSm,
  FradkinKivelsonPRB,LRDorseyQHN}.  The underlying unifying feature of
these diverse set of critical systems is the {\em spontaneously}
broken rotational (in contrast to e.g., density waves in crystalline
solids) and translational symmetries.  Based on these symmetries that
they break, we collectively refer to this class of states as smectics
akin to their conventional soft matter liquid crystal realizations,
that we turn to next.

\subsection{Smectic liquid crystals}

\begin{figure}[htbp]
\hspace{0in}\includegraphics*[width=0.7\textwidth]
{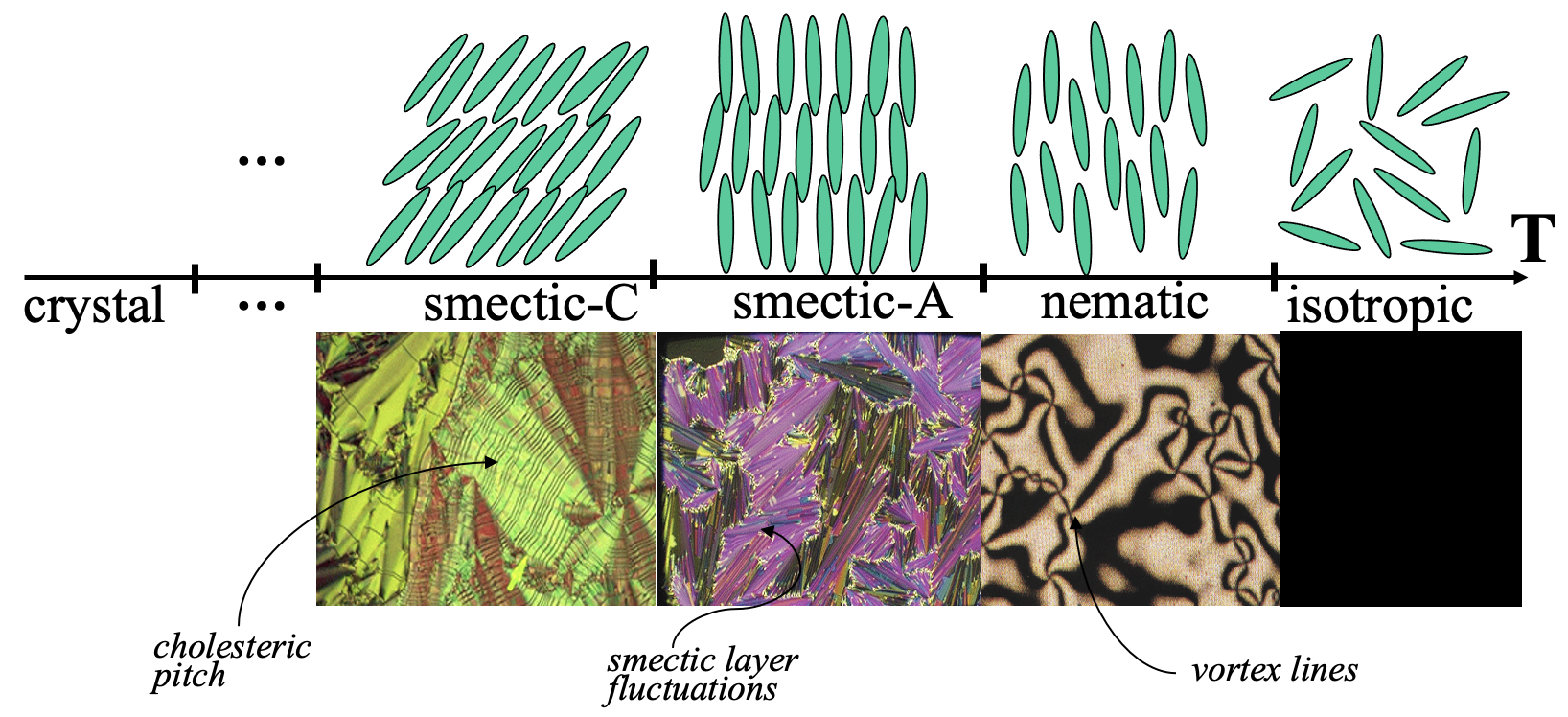}
\caption{Illustration of the most ubiquitous nematic (orientationally
  ordered uniaxial fluid), smectic-A and smectic-C (one-dimensional
  density wave with, respectively isotropic and polar in-plane fluid
  orders) liquid crystal phases and their associated textures in
  cross-polarized microscopy (N.A. Clark laboratory).}
\label{LCphasesFig}
\end{figure}

Illustrated in Fig.\ref{LCphasesFig}, the most ubiquitous liquid
crystal phases are the uniaxial nematic, that spontaneously breaks
rotational symmetry of the parent isotropic fluid and the smectic
state, a uniaxial one-dimensional density wave that further breaks
translational symmetry along a single axis. The uniaxial nematic is
characterized by a quadrupolar order parameter
$Q_{ij} = S(\nh_i\nh_j - \frac{1}{3}\delta_{ij})$, with $S$ the
strength of the orientational order along the principle uniaxial axis
$\hat{\bf n}$. A three-dimensional smectic-A is a periodic array of
two-dimensional fluids, characterized by a uniaxial periodic density
modulation with Fourier components that are integer multiples of the
smectic ordering wavevector ${\bf q_0}=\hat{{\bf n}} 2\pi/a $ (with
$a$ the layer spacing) parallel to the nematic director $\hat{\bf n}$;
other type of smectics, e.g., smectic-C, where $\hat{\bf n}$ makes a
nonzero angle with $\qv_0$ are common.\cite{ChaikinLubensky,deGennes}
The dominant lowest Fourier component $\psi(\xv)\equiv\rho_{\qv_0}$ can
be taken as the local (complex scalar) order parameter which
distinguishes the smectic-A from the nematic phase\cite{deGennes}. It
is related to the molecular density $\rho({\bf x})$ by
\begin{equation}
  \rho({\xv})= \mbox{Re}[\rho_0+\psi({\xv}) e^{i {\bf q_0}\cdot{\xv}}]\;,
\label{rho}
\end{equation}
where $\rho_0$ is the mean density of the smectic, $\psi=\rho_{\qv_0}$
is the Fourier amplitude of the density at wavevector $\qv_0$ and
$\tRe$ is a real part.

As first introduced by de Gennes\cite{deGennes}, the effective
Hamiltonian functional $H_{dG}[\psi,\hat{\bf n}]$, that describes the
nematic-to-smectic-A (NA) transition at long length scales, is given by,
\begin{eqnarray}
  H_{dG}[\psi,\hat{\bf n}]&=&\int \! d^dx
  \bigg[c|(\grad -
  i q_0 {\bf \delta\nh})\psi|^2
  + t_0|\psi|^2 +{1\over2} g_0 |\psi|^4\bigg]+H_F[\bf {n}]\;,
\label{HdG}
\end{eqnarray}
where $t_0\propto T-T_{NA}$ is the reduced temperature for transition
at $T_{NA}$,
\begin{equation}
  \delta\hat{\bf n}(\xv) \equiv \hat{\bf n}(\xv) - \hat{\bf n}_0 =
\delta\hat{\bf n}_\perp + \hat{\bf n}_0(\sqrt{1-\delta\hat{\bf
    n}_\perp^2}-1)
\end{equation}
is the fluctuation of the local nematic director $\hat{\bf n}(\xv)$
away from its average value $\hat{\bf n}_0$, which I take to be
$\hat{\bf z}$, and $H_F[\bf n]$ is the Frank effective Hamiltonian
that describes the elasticity of the nematic director:
\begin{eqnarray}
H_{F}[\hat{\bf n}] = \int \! d^d x\;
{1\over 2} \bigg[K_s(\grad\cdot\hat{\bf n})^2+
K_t(\hat{\bf n} \cdot \grad\times\hat{\bf n})^2
 + K_b(\hat{\bf n}\times\grad\times\hat{\bf n})^2\bigg]\;,
\label{Frank}
\end{eqnarray}
where $K_s$, $K_t$, and $K_b$ are the bare elastic moduli for splay,
twist and bend of the nematic director field,
respectively.\cite{deGennes, ChaikinLubensky}

The ``minimal'' gauge-like coupling between $\bf n$ and $\psi$ is
enforced by the requirement of global rotational
invariance\cite{deGennes}.  It is important to emphasize, however,
that although de Gennes Hamiltonian $H_{dG}$ closely resembles the
Ginzburg-Landau model of a superconductor and the Abelian-Higgs model,
there are essential differences, that are discussed in the chapter by
Tom Lubensky. The physical reality of the nematic director ${\bf\nh}$
and the smectic order parameter $\psi$ (in contrast to the gauge
ambiguity in the definition of the vector potential and the
superconducting order parameter), selects the liquid crystal gauge
${\delta\bf\nh}_\perp\cdot{\bf\nh}_0=0$ as the physical gauge in which
$\bf\nh$ and $\psi$ are measured. The strict gauge invariance is also
explicitly broken in $H_{dG}$ by the splay term
$K_s(\grad\cdot\hat{\bf n})^2$ of the Frank Hamiltonian (contrasting
with the Maxwell action that involves purely gauge invariant
derivatives, e.g., $(\grad\times\Av)^2$), that, as we will see below
determines the smectic Goldstone-mode elasticity.

\subsection{Smectic elasticity}

\subsubsection{Smectic from a nematic liquid crystal fluid}

Within the ordered smectic phase, the fluctuations are conveniently
described in terms of the magnitude and phase of the smectic order
parameter $\psi$. It is easy to show that the fluctuations of the {\em
  magnitude} of $\psi$ around the average value
$|\psi_0|=\sqrt{t_0/g_0}=\mbox{const.}$ are ``massive'', and can
therefore be safely integrated out of the partition function, leading
to only {\em finite}, unimportant shifts in the effective elastic
moduli. In contrast, the phase of $\psi$ is a U(1) massless Goldstone
mode, corresponding to spontaneously broken translational symmetry
along ${\bf q}_0$. It is the low-energy phonon degree of freedom of
the smectic state, describing local displacement of the smectic layers
from its perfect periodic order. In accord with this discussion, deep
within the smectic phase, we can represent the smectic order parameter
as
\begin{eqnarray}
  \psi(\xv)=|\psi_0|e^{-i q_0 u(\xv)}\;,
  \label{psiSm}
\end{eqnarray}
safely neglecting fluctuations in the magnitude $|\psi_0|$ at
temperature below the $T_{NA}$.

Using this low-temperature smectic description \rf{psiSm} inside the
de Gennes Hamiltonian \rf{HdG} and dropping constants, one finds
\begin{eqnarray}
  H[u,\delta{\bf\nh}_\perp]
&=&\int\! d^dx \bigg[{B\over 2}
(\grad_\perp u+\delta{\bf\nh}_\perp)^2
+{B\over 2}(\partial_z u - \oh\delta\nh_\perp^2)^2\;\nonumber\\
&&+{K_s\over2}(\grad\cdot\delta{\bf\nh})^2
+{K_t\over2}(\hat{\bf z}\cdot\grad\times\delta{\bf\nh})^2
+ {K_b\over2}(\hat{\bf z}\times\grad\times\delta{\bf\nh})^2\bigg],
\label{Hun}
\end{eqnarray}
where $B=2c|\psi_0|^2 q_0^2$ is the smectic compression modulus. I
observe that the fluctuation mode $\grad_\perp u+\delta{\bf\nh}_\perp$
is ``massive'' and leads to an emergent Anderson-Higgs-like mechanism,
a hallmark of gauge theories.  As a consequence, after a simple
Gaussian integration over $\delta {\bf\nh}_\perp$, one finds that at
long length scales, $\delta{\bf\nh}_\perp$ fluctuations are
constrained to follow $\grad_\perp u$, corresponding to locking of the
director $\hat{\bf n}$ to the smectic layer normal. The low-energy
elastic smectic Hamiltonian is then obtained by the replacement
\begin{equation}
\delta{\bf\nh}_\perp\rightarrow -\grad_\perp u\;,
\end{equation}
everywhere in Eq.\rf{Hun} and Frank energy Eq.\rf{Frank}. Valid in the
long wavelength limit and provided dislocations are confined, one thus
obtain a nonlinear elastic Goldstone-mode Hamiltonian of the smectic
phase,
\begin{eqnarray}
H_{sm}[{u}]&=&\int \! d^d x \bigg[{K\over 2}(\nabla^2_\perp {u})^2
+ {B\over 2}(\partial_z {u} - \oh(\grad u)^2)^2\bigg]\;,
\label{Hu_el}
\end{eqnarray}
where $K=K_s$ is the splay modulus.\cite{ChaikinLubensky,GP}

I observe that the harmonic elasticity of the smectic --
$m=d-1$-Lifshitz model\cite{mLifshitz} -- is highly anisotropic (at
harmonic level with scaling $z\sim x_\perp^2$), displaying compression
modulus $B$ along ${\bf q}_0$ and higher-order {\em curvature}
(Laplacian) modulus transverse to ${\bf q}_0$ , with the ``tension''
(transverse gradient) modulus vanishing exactly. This is ensured by
the underlying rotational symmetry, as discussed in the Introduction
to critical phases.  As I will show in forthcoming sections, as a
consequence, smectic fluctuations are highly enhanced (relative to
e.g., its XY model counterpart), resulting in importance of nonlinear
Goldstone-modes' elasticity for $d\le 3$, that I therefore retained in
\rf{Hu_el}.  I also note that the compressional modulus $B$ multiplies
a square of the nonlinear strain tensor, whose precise form is
determined by the underlying invariance under arbitrary large rotation
of smectic layers,
\begin{eqnarray}
  u_0(\rv)=z(1-\cos\theta) + x\sin\theta,
\label{theta0rotate}
\end{eqnarray}
and maintained under coarse-graining RG.  I conclude this subsection
by also noting that the aforementioned connection to the
Anderson-Higgs mechanism elucidates why a smectic state is
characterized by only a single Goldstone mode $u$, rather than three,
despite partially breaking two rotational and one translational
symmetries. In contrast, a charged superconductor is well-known to
have all its Goldstone modes ``eaten'' -- fully gapped out by the
Anderson-Higgs mechanism.

\subsubsection{Smectic from an isotropic fluid}

I now derive the nonlinear smectic elasticity in a more basic,
complementary way, starting instead with an isotropic fluid
state.\cite{GP} I begin with a generic energy functional that captures
system's tendency to develop a unidirectional wave at wavevector
$\qv_0$, with an arbitrary direction, and magnitude fixed at $q_0$,
\begin{equation}
\cH_{sm} = \oh J\left[(\nabla^2\rho)^2 - 2q_0^2(\nabla\rho)^2\right] +
\oh t\rho^2 - w\rho^3 + v\rho^4 + \ldots,
\label{cHsm} 
\end{equation}
where $J, q_0, t, w, v$ are parameters of the isotropic fluid
phase. Clearly, the first term is engineered so that dominant
fluctuations and condensation are on a spherical surface at a nonzero
wavevector with a magnitude $q_0$. Thus I focus on the density at a
wavevector $\qv$,
that for now is unrelated to $q_0$
\begin{equation}
  \rho(\xv) = \tRe\left[\rho_q(\xv) e^{i\qv\cdot\xv}\right],
  \label{rhoSm}
\end{equation}
where $\rho_q(\xv)$ is a complex scalar.  Without loss of generality,
based on the discussion of the previous subsection, the order
parameter $\rho_q(\xv) = |\rho_q| e^{-i q u}$ is taken to have a
(constant) magnitude $|\rho_q|$ and a phase $q u(\xv)$.
Clearly $u(\xv)$ is just a phonon displacement along $\qv$.
Substituting this expression for $\rho(\xv)$ and its gradients into
$\cH_{sm}$ one finds
%
\begin{eqnarray}
  \cH_{sm}
&=& J\rho_0^2\left[\frac{1}{4}q^2(\nabla^2 u)^2 
+\bigg(q\qv\cdot\nabla u-\oh q^2(\nabla u)^2\bigg)^2
+4(q^2-q_0^2)\bigg(q\qv\cdot\nabla u
-\oh q^2(\nabla u)^2\bigg)\right]
    + \ldots,
 \label{Helast3}
\end{eqnarray}
%
where constant parts as well as fast oscillating pieces were dropped
as they average away after spatial integration of the above energy
density.  Firstly, I observe that (as discussed on general grounds
above) in a harmonic part linear gradient elasticity in $u$ only
appears for gradients {\em along} $\qv$, namely $\qv\cdot\nabla$, with
elasticity transverse to $\qv$ starting with a Laplacian
form. Secondly, the elastic energy density is an expansion in a
rotationally-invariant strain tensor combination
\begin{equation}
u_{qq}=\hat{\qv}\cdot\nabla u -\oh(\nabla u)^2,
\label{uqq}
\end{equation}
whose nonlinearities in $u$ ensure that it is fully rotationally
invariant even for large rotations. To see this (picking
$\hat\qv ={\bf\zh}$) note that a rigid (distortion-free) rotation of
$\qv$
($q_0{\bf\zh}\rightarrow\qv = q_0(\cos\theta{\bf \zh} +
\sin\theta{\bf\xh}$), can be interpreted as a spatially linear
``distortion'' $u_0(\xv)=z(1-\cos\theta) + x\sin\theta$, for which
nonlinear strain $u_{qq}$ vanishes identically, thus, as required,
corresponds to a vanishing energy. Thirdly, the last term in
\rf{Helast3} vanishes for $|\qv|$ picked to equal $q_0$, corresponding
to energetically preferred choice of modulation wavevector.

Looking ahead, as one includes effects of fluctuations, the ``bare''
condition $q=q_0$ will need to be adjusted so as to eliminate the
fluctuation-generated linear term in $u_{qq}$ order by order, which
amounts to an expansion in the nonlinear strain $u_{qq}$ around the
correct (fluctuation-corrected) ground state. Finally, I note that the
relation between the curvature modulus $K$ of Laplacian (first) term
and the bulk modulus $B$ gradient (second) term is not generic and can
be relaxed to have distinct elastic constants, as can be seen if
higher order gradient terms are included in the original energy
density, \rfs{cHsm}.

Choosing the coordinate system such that $\hat{\bf z}$ is aligned
along $\qv$, one finds that for $q=q_0$, the Goldstone-mode
Hamiltonian \rf{Helast3} reduces to the standard smectic elastic
energy density, \rf{Hu_el} derived from the de Gennes model
above. \cite{deGennes,ChaikinLubensky}.

\subsubsection{Smectic as cholesteric and helical fluids}

Another beautiful example of a system that realizes a uniaxial
periodic critical state is a cholesteric phase of chiral liquid
crystals, illustrated in Fig.\ref{criticalPhasesFig}(b). It is in fact
the first liquid crystal phase discovered in cholesterol benzoate by
Reinitzer in 1886.\cite{deGennes, ChaikinLubensky} Such orientational
(as opposed to mass) density wave state is ubiquitous in nature and is
equivalent to the helical co-planar spin-spiral state that is found in
non-centrosymmetric magnets like MnSi, FeSi and many others\cite{MnSi}
(neglecting crystal-field pinning that is always present due to
spin-orbit interactions and explicit breaking of rotational
symmetry.).

Following analysis of Radzihovsky and Lubensky\cite{cholestericLR},
below I discuss the low-energy {\em nonlinear} elasticity of a
cholesteric, on length scales longer than its period, $a = 2\pi/q_0$.
As first proposed by de Gennes, the latter is expected, based on
symmetry and an explicit {\em harmonic} derivation by
Lubensky\cite{cholestericLubensky,deGennes}, to be identical to that
of a smectic, i.e., that of a one-dimensional crystalline order, that
spontaneously breaks underlying translational and rotational symmetry,
with constant-orientation ``layers'' transverse to the ordering
wavevector, ${\qv}_0$. Although the underlying chirality has only
subtle surface effects within the cholesteric state, its effects
become important in the nature of the phase transition in and out of
the cholesteric critical state.

The simplest model of a chiral nematic liquid crystal (a cholesteric),
is captured by the chiral Frank elastic free energy density,
\begin{eqnarray}
\cH^*_F&=&\oh K_s(\nabla\cdot\nh)^2 + \oh K_b(\nh\times\nabla\times\nh)^2 
           + \oh K_t(\nh\cdot\nabla\times\nh+q_0)^2.
\label{H*_F}
\end{eqnarray}
The broken chiral symmetry allows a chiral $q_0$ term, that leads to a
twist of the nematic structure into a helix with a pitch $2\pi/q_0$
along a spontaneously chosen axis.


In a simplified isotropic limit of $K_s=K_b=K_t$, the chiral Frank
energy density is then given by
\begin{eqnarray}
\cH^*_F&=&\oh K\left[(\partial_i\nh_j)^2+2q_0\nh\cdot\nabla\times\nh +
  q_0^2\right],\\
&=&\oh K\left(\partial_i\nh_j+q_0\epsilon_{ijk}\nh_k\right)^2
- \oh K q_0^2,
\label{Hi*_F}
\end{eqnarray}
where after integration by parts I utilized the identity
\begin{eqnarray}
(\partial_i\nh_j)^2&=&(\nabla\cdot\nh)^2
+ (\nabla\times\nh)^2 + \nabla\cdot\big[(\nh\cdot\nabla)\nh -
\nh\nabla\cdot\nh\big].
\label{24identity}
\end{eqnarray}

In the absence of topological defects, neglecting the boundary term,
$\cH^*_F$ is a sum of squares and is therefore minimized by a
twist-only cholesteric state
\begin{eqnarray}
  \nh(\rv)&=&\eh_{10}\cos(\qv\cdot\rv)+\eh_{20}\sin(\qv\cdot\rv),
\end{eqnarray}
where $\eh_{10},\eh_{20}, \eh_{30}\equiv\eh_{10}\times\eh_{20}$ form
an orthonormal triad, with a constant twist
\begin{equation} 
\nh\cdot\nabla\times\nh = -q_0.
\end{equation}



I now derive the low-energy Goldstone-mode elasticity about the
cholesteric helical ground state, that can be parameterized according
to
\begin{eqnarray}
  \nh(\rv)&=&\eh_1(\rv)\cos(\qv\cdot\rv+\chi(\rv))
+\eh_2(\rv)\sin(\qv\cdot\rv+\chi(\rv)),
\label{nGeneral}
\end{eqnarray}
where fluctuations are captured by the spatially dependent orthonormal
triad $\eh_1(\rv),\eh_2(\rv), \eh_3(\rv)$ and the helical phase
\begin{equation}
  \chi(\rv)=-q_0 u(\rv),
\end{equation}
that corresponds to the phonon field $u(\rv)$ of the
constant-orientation layers.

The helical state breaks a group of three dimensional translations and
rotations $G=T_{x,y,z}\times O(3)$ of the isotropic fluid down to
$H = T_x\times T_y\times U(1)$ (latter
$U(1) = \text {diagonal}[T_z,O_z(2)]$). Thus, since
dim$[G/H = O(3)]=3$, one may expect {\em three independent} Goldstone
modes $\chi(\rv)$ and $\eh_3(\rv)$, corresponding to three degrees of
freedom of the orthonormal triad; the azimuthal angle $\phi(\rv)$
defining the orientation of the $\eh_{1,2}$ around $\eh_3$ is not
independent of $\chi(\rv)$ as it can be absorbed into it. The
low-energy coset space is isomorphic to $S^1\times S^2$, a ball
(radius $\pi$) of the group manifold of SO(3).  However, despite of
this standard counting, as I will show below and is expected on
general grounds anticipated by de Gennes, the cholesteric state is
characterized by a {\em single} smectic-like Goldstone mode $u(\rv)$.

To this end, substituting the form for ${\bf\nh}(\rv)$ from
\rfs{nGeneral} into the Frank energy of the chiral nematic \rf{H*_F},
and defining effective connection gauge fields, ${\bf a}$ and
${\bf c}_{1,2}$ \bse
\begin{eqnarray}
\partial_i\eh_{1j}&=&a_i\eh_{2j} + c_{1i}\eh_{3j},\\
\partial_i\eh_{2j}&=&-a_i\eh_{2j} + c_{2i}\eh_{3j},\\
a_i=\eh_{2}\cdot\partial_i\eh_{1},\ \
c_{1i}&=&-\eh_{1}\cdot\partial_i\eh_{3},\ \
c_{2i}=-\eh_{2}\cdot\partial_i\eh_{3},
\label{identies}
\end{eqnarray}
\ese
%
one finds,
\begin{eqnarray}
  \cH^*_F&=&
             \frac{K}{2} (\nabla\chi + \av + \qv - q_0\eh_{3})^2
+ \frac{K}{4} (\cv_1 + q_0\eh_{2})^2
    + \frac{K}{4} (\cv_2 - q_0\eh_{1})^2.
\label{Hchi}
\end{eqnarray}
In above, I again neglected spatially oscillating and constant terms.

Taking $\qv=q_0{\bf \zh}$, with ${\bf \zh}$ defining the helical axis
(distinct from the normal to the helical plane, $\eh_3$) and noting
the compatibility condition on effective flux or its equivalent
vanishing of the Pontryagin density
\begin{eqnarray}
\nabla\times\av
=\epsilon_{ij}\eh_3\cdot\partial_i\eh_3\times\partial_j\eh_3=0,
\end{eqnarray} 
required by well-defined cholesteric layers, i.e., in the absence of
dislocations and disclinations in the layer structure, allows one to
take
\begin{eqnarray}
\av=\grad\phi.
\end{eqnarray}
Under this condition $\phi$ can be eliminated in favor of $\chi$,
i.e., $\chi+\phi\rightarrow\chi$ and in the laboratory coordinate
system ${\bf\xh},{\bf\yh},{\bf\zh}$, the fluctuations are
characterized by the local helical frame described by $\chi$ and
$\eh_3$, with
\begin{eqnarray}
\eh_3&=&\eh_{3\perp}+{\bf\zh}\sqrt{1-\eh_{3\perp}^2}
\approx\eh_{3\perp}+{\bf\zh}(1-\oh\eh_{3\perp}^2).
\end{eqnarray}
I thus obtain
\begin{eqnarray}
\cH^*_F&=&\oh (\nabla_\perp\chi - q_0\eh_{3\perp})^2
+ \oh (\partial_z\chi + \oh q_0\eh_{3\perp}^2)^2
+ \frac{1}{4} (\cv_1^2 + \cv_2^2)
+ \frac{q_0}{2} (\cv_1\cdot\eh_{2} - \cv_2\cdot\eh_{1}).
\label{Hchi2}
\end{eqnarray}
From the minimization of the first term (or equivalently integrating
out the independent $\eh_{3\perp}$ degree of freedom), I obtain an
effective constraint
\begin{equation}
\nabla_\perp\chi = q_0\eh_{3\perp},
\end{equation}
that is an example of an emergent Higgs mechanism (akin to smectic
liquid crystals discussed above) locking the cholesteric layers with
the molecular frame orientation. With this constraint (valid at low
energies) the effective cholesteric Hamiltonian reduces to
\bse
\begin{eqnarray}
\cH^*_F&=&\oh\big[\partial_z\chi + \frac{1}{2q_0}(\nabla_\perp\chi)^2\big]^2
+ \frac{1}{4q_0^2}(\eh_{\alpha\gamma}\cdot
\partial^\perp_\beta\partial^\perp_\gamma\chi)^2
+ \frac{q_0}{2}
(\eh_{1\alpha}\eh_{2\beta}\partial_\beta\partial_\alpha\chi
-\eh_{2\alpha}\eh_{1\beta}\partial_\beta\partial_\alpha\chi),\hspace{0.5cm}\\
       &=&\frac{B}{2}\big[\partial_z u - \oh(\nabla u)^2\big]^2
           + \frac{K}{2}(\nabla^2 u)^2
\label{Hchi3}
\end{eqnarray}
\ese
where compressional modulus $B= K q_0^2$,
$\eh_{\alpha, i}\eh_{\alpha, j} = \delta_{ij}$ and
$\eh_{1, i}\eh_{2, j} - \eh_{2, i}\eh_{i, j} =
\eh_{3,k}\epsilon_{ijk}$ were used, and to eliminate the last term in
the first line above I used the condition of the single-valuedness of
the phase field $\chi$, i.e., well defined cholesteric layers with no
dislocations.  However, this last condition is violated and thus
distinguishes the cholesteric from a smectic near a phase transition
out of the cholesteric state.

Thus, as advertised and expected on symmetry grounds one indeed
finds\cite{cholestericLR}, that cholesteric Goldstone-mode elasticity,
even at the nonlinear level is identical to that of a conventional
smectic -- at harmonic level controlled by the $m=d-1$-Lifshitz
model\cite{mLifshitz}.  Thus smectic critical state universal
phenomenology, to be worked out below, will equally apply to the
cholesteric state.

 
\subsubsection{Smectic as a helical bosonic superfluid}

I next discuss a quantum realization of a smectic state, namely via
interacting superfluid bosons on a frustrated honeycomb
lattice.\cite{helicalBosonsHMR22} Other bosonic realizations
also appear in p-wave Feshbach-resonant\cite{LR_ChoiPRL} and Rashba
spin-orbital coupled \cite{HuiZhai} bosons, engineered in a controlled
way in cold atom experiments\cite{MonikaAidelsburger, HuiZhai} to
condense at a nonzero momentum, ${\bf k}_0$.

\begin{figure}[h]
  \hspace{-1cm}
  \includegraphics[width=0.28\textwidth]{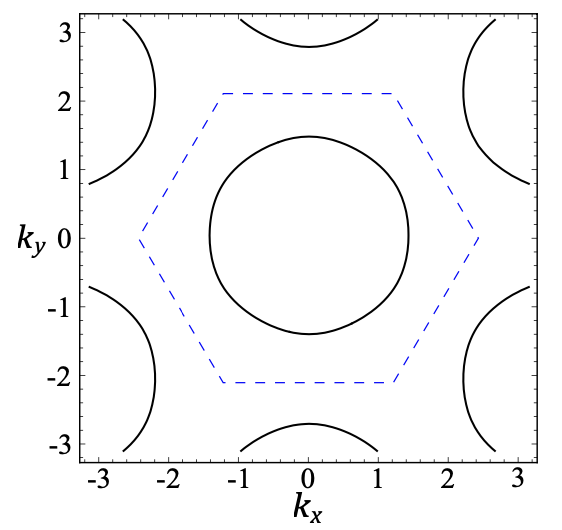}
  \hspace{1cm}
  \includegraphics[width=0.4\textwidth]{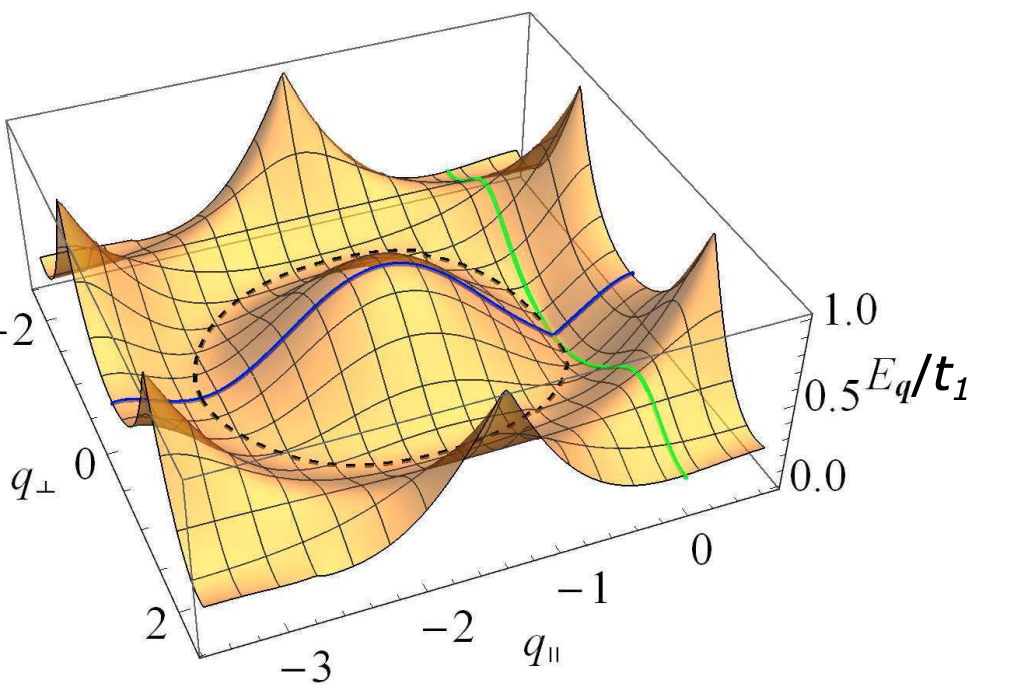}
  \put(-45,85){ \includegraphics[width=0.15\textwidth]{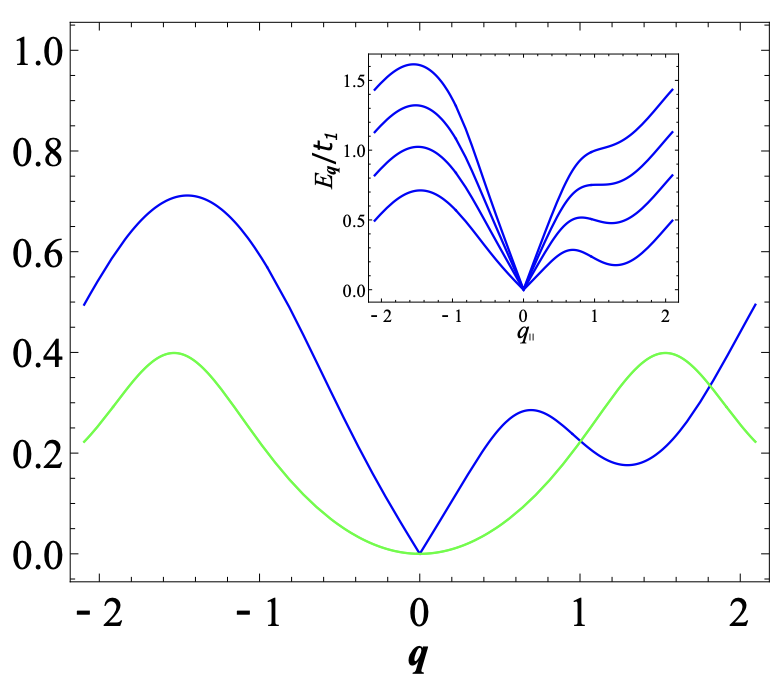}}
  \caption{(a) A black circular curve is a nonzero momentum $\kv_0$
    minimum in the noninteracting microscopic dispersion
    $\epsilon^-_\kv$ (around the $\Gamma$ point in the Brillouin zone,
    dashed blue curve) of bosons on a frustrated honeycomb lattice,
    \rf{Heq}.  (b) The dispersion of the interacting helical
    superfluid, $E_{\qv}$ around ${\bf q}=\kv - \kv_0=0$, with green
    (see inset) indicating quadratic $q_\perp^2$ form along $q_\perp$
    and blue (see inset) $|q_\parallel|$ along the condensate
    momentum, $\kv_0$, consistent with symmetry-expected smectic form,
    given by \rf{eq:L_phi_0}.  The black-dashed contour indicates the
    degenerate minimum of the noninteracting band
    $\epsilon^-_{\bf k}$.}
\label{dispersionk0}
\end{figure}

I refer the reader to the original literature for these latter systems
and here focus on bosons on a frustrated honeycomb latice, modeled by
a tight-binding dispersion with nearest $t_1$ and next-nearest
(antiferromagnetic) $-t_2$ neighbor hopping amplitudes, with
\begin{equation}
H = -t_1\sum_{\langle ij\rangle}  a^\dag_{i,1}
a_{j,2}+t_2\sum_{\langle \langle ij \rangle\rangle}  (a^\dag_{i,1}
a_{j,1}+a^\dag_{i,2} a_{j,2} )+h.c.
+\frac{U}{2}\sum_{i}\sum_{s=1,2} n_{i,s}(n_{i,s}-1).
\label{Heq}
\end{equation}
For sufficiently frustrated hopping, with $t_2/t_1> 1/6$ (that has
been engineered in an optical lattice through Floquet techniques
\cite{MonikaAidelsburger}), above system displays a dispersion minimum
at a nonzero momentum closed contour, illustrated in
Fig.\ref{dispersionk0}.

Such nonzero dispersion, with a closed minimum contour is
approximately circular for $t_2/t_1 \approx 1/6 +\epsilon$ and can be
faithfully modeled by a continuum field theory, encoded in a
noninteracting Hamiltonian
$H_0=\int_{\bf r}\hat{\Psi}_{\bf r}^\dag\hat{\varepsilon}_{{\bf
    r}}\hat{\Psi}_{\bf r}$ with,
\begin{equation}
  \hat{\varepsilon}_{{\bf r}} = J(-\nabla^2-{k}_0^2)^2+\varepsilon_0 ,
  \label{eq:eff_th_free_dispersion}
\end{equation}
with a quartic dispersion
\begin{equation}
  \varepsilon_{{\bf k}} = J({\bf k}^2-{k}_0^2)^2 + \varepsilon_0 ,
 \label{eq:dispersion_isotropic}
\end{equation}
and minima lying on contour ${\bf k}^2={k}_0^2$.  The
corresponding Euclidean (imaginary-time $\tau$) Lagrangian is given
by,
\begin{eqnarray}
  \mathcal{L} &=&  \Psi^\ast \partial_\tau \Psi +J | \nabla^2 \Psi|^2
                  -2J{k}_0^2 |\nabla \Psi|^2 + (\tilde{\varepsilon}_0-\mu)|\Psi |^2
                  +\frac{U}{2} |\Psi |^4 .
\label{Lpsi}
\end{eqnarray}

For $\mu > \epsilon_0$ bosons condense into a nonzero-momentum
superfluid state, a simplest version of which is a helical condensate
at a single point ${\bf k}_0$ on the dispersion contour minimum --
encoded in $\varepsilon_{\bf k}$, (\ref{eq:dispersion_isotropic}). In
the density-phase representation, the state is characterized by
\begin{equation}
\Psi ({\bf r}) = \sqrt{n} e^{i {\bf k}_0 \cdot {\bf r} + i\phi} = \sqrt{n_0 +\pi} e^{i {\bf k}_0 \cdot {\bf r} + i\phi} \label{eq:psi_density_phase}
\end{equation}
with a nonzero condensate density (at mean-field level)
$n_0=(\mu-\varepsilon_0)/U$ and momentum $k_0$, for
$\mu>\varepsilon_0$ (vanishing otherwise), and density and phase
fluctuations, $\pi$ and $\phi$, respectively. To quadratic order, the
helical superfluid is then described by a Goldstone-mode Lagrangian
density, which, after integrating out density field $\pi$ gives a
zero-temperature quantum ``columnar'' Lagrangian -- $m=d-2$-Lifshitz
model in $d$ space-time dimensions\cite{mLifshitz} -- (to also appear
in subsequent sections)
\begin{align}
  \mathcal{L}_\phi
    \approx &~ B_\tau (\partial_\tau \phi)^2 +  B (
              \partial_\parallel \phi )^2 +  K (\partial_\perp^2
              \phi)^2.
              \label{eq:L_phi_0}
\end{align}
At nonzero $T$, \rf{eq:L_phi_0} reduces to the classical smectic
($m=d-1$-Lifshitz model\cite{mLifshitz}) Hamiltonian for $\phi$
discussed in previous subsections.

\subsubsection{Smectic as a Fulde-Ferrell-Larkin-Ovchinnikov superconductor}

A singlet superconductor, frustrated by a depairing Zeeman field has
been predicted by Fulde and Ferrell\cite{FF} and by Larkin and
Ovchinnikov\cite{LO} (FFLO) to pair at a nonzero momentum, thereby
exhibiting a periodic modulation of the Cooper-pair amplitude -- a
pair-density wave (PDW)\cite{ffloLR,PDWaop}. Such FFLO state, in the
absence of deleterious orbital field and pinning lattice effects was
proposed\cite{SRprl,SRaop,SRreview} to be realizable in spin
imbalanced (polarized) Feshbach-resonant paired atomic superfluids. It
is now believed to have been observed in an array of decoupled
one-dimensional traps, with recent progress toward a higher
dimensional realization of the FFLO superfluid.\cite{Hulet}

While microscopic analysis is a bit involved\cite{SRprl,SRaop,ffloLR},
its upshot is a derivation of an effective Landau theory of the form
\rf{Lpsi} for the pairing amplitude, with a harmonic disperson
$\varepsilon_k$, for intermediate Zeeman field (chemical potential for
spin imbalance) $h_{c1} < h < h_{c2}$, displaying a minimum on nonzero
momentum closed contour, akin to that of the helical superfluid
\rf{Lpsi}, above.  Fulde-Ferrell and Larkin-Ovchinnikov are two
simplest states that minimize this Lagrangian. The FF state is
characterized by a pairing amplitude, \cite{FF}
\begin{eqnarray}
\Psi_{FF}(\rv)=\Psi_{q_0} e^{i\qv_0\cdot\rv +i\phi},
\end{eqnarray}
that is a plane-wave with the momentum $\qv_0$ and a single Goldstone
mode $\phi$,
a local superconducting phase. The state carries a nonzero, uniform
spontaneously-directed supercurrent
\begin{eqnarray}
  \jv_{FF} = \frac{1}{m}|\Psi_{q_0}|^2(\qv_0+\nabla\phi),
\label{jFF}
\end{eqnarray}
and thereby breaks time-reversal and rotational symmetries, chosen
spontaneously along $\qv_0$, as well as the global gauge symmetry,
corresponding to the total atom conservation.  The low-energy
Lagrangian for $\phi(\rv,\tau)$ takes the smectic (columnar at $T=0$)
form, \rf{eq:L_phi_0}.

The LO state\cite{LO} is a time-reversal symmetric counterpart, that
is a superposition of $\pm {\bf q}_0$ paired condensates, given by,
\bse 
\begin{eqnarray}
\Psi_{LO}(\rv) &=&\Psi_+(\rv) e^{i\qv\cdot\rv} + 
                     \Psi_-(\rv) e^{-i\qv\cdot\rv},\\
&=&2|\Psi_{q_0}|e^{i\oh(\phi_+ + \phi_-)}
\cos\big[\qv_0\cdot\rv + \oh(\phi_+ - \phi_-)\big],\ \ \ \ \ \ \ \\
&=&2|\Psi_{q_0}|e^{i\phi}\cos\big[\qv_0\cdot\rv + \theta\big].
\end{eqnarray}
\label{DeltaLO}
\ese
It is characterized by the superfluid phase $\phi$ and the phonon
$u = \theta/q_0$ Goldstone modes, with a combination of the XY and
smectic sectors in a classical Hamiltonian,
\begin{eqnarray}
  \cH_{LO} &=&\oh K(\nabla^2 u)^2 + 
\oh B\big(\partial_\parallel u - \frac{1}{2}(\nabla u)^2\big)^2
+ \frac{1}{2}\rho_s^\parallel(\partial_\parallel\phi)^2 
+ \frac{1}{2}\rho_s^\perp(\nabla_\perp\phi)^2,
\label{HgmLO}
\end{eqnarray}
whose zero-temperature quantum counterpart is a combination of XY and
``columnar'' -- $m=d-2$-Lifshitz model in $d$ space-time
dimensions\cite{mLifshitz} forms.  At nonzero temperature one thus
expects FF and LO states to display critical smectic phenomenology
that I discuss in the next section.

\subsubsection{Smectic as quantum Hall stripes}

One other prominent example of a quantum smectic has been argued
theoretically\cite{Fogler,Moessner,MacDonaldFisherSm,FradkinKivelsonPRB}
to be realized in half-filled large $N$ Landau level regime and
observed experimentally\cite{EisensteinSm, Du99} at such filling
$\nu = N+1/2$ through appearance of a highly anisotropic transport
below 100 milli-Kelvin.  At mean-field level the state is described as
a periodic array of alternative integer filling $N$ and $N+1$ quantum
Hall stripes, predicted to be the exact ground state in the $N \gg 1$
limit. Beyond Hartree-Fock theory\cite{Fogler,Moessner}, fluctuations
of such a striped state are well described by chiral edge Luttinger
liquids, corresponding to inter- (positions) and intra- (shape) stripe
phonons of this quantum Hall smectic.  Much of the analysis discussed
below, importantly beyond its harmonic fluctuations, will also apply
to this state at nonzero temperature leading to a critical smectic
state. At zero temperature, it is described by ``cholesteric''-like,
i.e., $m=d-2$-Lifshitz model in $d$ space-time
dimensions\cite{mLifshitz}.

\subsection{Nonzero $T$ Gaussian fluctuation in a harmonic smectic}

To assess the extent of thermal fluctuations of the smectic Goldstone
mode $u(\rv)$ I first analyze them within a harmonic approximation,
neglecting elastic nonlinearities in $H_{sm}$.
\cite{ChaikinLubensky,Caille} In terms of the Fourier modes $u_\kv$,
the Hamiltonian decouples, reducing to
\begin{eqnarray}
H_{sm} = \oh \int \frac{d^d k}{(2\pi)^d} 
\left(K k_\perp^4 + B k_z^2\right)|u_\kv|^2,
\label{Hsm_k}
\end{eqnarray}
thus allowing a straightforward computation of phonon correlation
functions via standard Gaussian integrals or equivalently via
equipartition. This gives mean-squared fluctuations
%
\begin{eqnarray}
\langle u^2\rangle_0^{T}
&=&\int^{\Lambda_\perp}_{L_\perp^{-1}}\frac{d^dk}{(2\pi)^{d}}
\frac{T}
{B k_z^2 + K k_\perp^4} 
\approx
\left\{\begin{array}{ll}
\frac{T}{2\sqrt{B K}}C_{d-1}L_\perp^{3-d},& d < 3,\\
\frac{T}{4\pi\sqrt{B K}}\ln q_0L_\perp,& d = 3,\\
\end{array}\right.
\label{uuT}
\end{eqnarray}
%
%
%
where I defined a constant
$C_d=S_d/(2\pi)^d=2\pi^{d/2}/[(2\pi)^d\Gamma(d/2)]$, with $S_d$ a
surface area of a $d$-dimensional sphere, and introduced an infrared
cutoff by considering a system of finite extent $L_\perp\times L_z$,
with $L_z$ the length of the system along the ordering ($z$) axis and
$L_\perp$ transverse to $z$.  Unless it has a huge aspect ratio, such
that $L_z \sim L_\perp^2/\lambda>> L_\perp$, any large system
($L_\perp,L_z >>\lambda$) will have $\lambda L_z \ll L_\perp^2$.

The key observation here is that the smectic phonons exhibit
fluctuations that diverge, growing logarithmically in 3d and linearly
in 2d with system size $L_\perp$; for $d > 3$ fluctuations are
bounded. Thus, as a consequence of their ``soft'' elasticity (a
vanishing $(\grad_\perp u)^2$ modulus) 3d smectics are akin to 2d XY
systems, such as superfluid films and two-dimensional crystals,
\cite{Landau,Peierls,MerminWagner,Hohenberg,Coleman}, exhibit a
power-law order in three dimensions.

The expression for the mean-squared phonon fluctuations in \rf{uuT}
leads the emergence of important crossover length scales
$\xi_\perp,\xi_z$, related by
%
\begin{eqnarray}
  \xi_\perp & = & (\xi_z\sqrt{K/B})^{1/2}\equiv\sqrt{\xi_z\lambda},
\label{xi_pxi_z}
\end{eqnarray}
%
that characterize the finite-temperature smectic state. These are
defined as scales $L_\perp,L_z$ at which phonon fluctuations are
large, comparable to the smectic period $a=2\pi/q_0$. Namely, setting
\begin{equation}
\langle u^2\rangle_0^{T}\approx a^2
\end{equation}
in \rfs{uuT} one finds
%
\begin{eqnarray}
\xi_\perp&\approx&
\left\{\begin{array}{ll}
\frac{a^2\sqrt{B K}}{T}\sim\frac{K}{T q_0},& d = 2,\\
a e^{4\pi a^2\sqrt{B K}/T}\sim a e^{\frac{c K}{T q_0}},& d = 3,\\
\end{array}\right.
\label{xiperp}
\end{eqnarray}
%
where in the second form of the above expressions I took the simplest
approximation for the smectic anisotropy length $\lambda=\sqrt{K/B}$
to be $\lambda = a \sim 1/q_0$, and introduced $O(1)$ Lindemann
constant $c$,\cite{Lindemann} that depends on the somewhat arbitrary
definition of ``large'' phonon root-mean-squared fluctuations.

The smectic connected correlation function 
\begin{equation}
C_{u}(\xv_\perp,z)
=\langle\left[u({\xv_\perp},z)-u({\bf 0},0)\right]^2\rangle_0\;.
\label{C_T}
\end{equation}
is also straightforwardly worked out, in 3d giving the logarithmic
Caill\'e form\cite{Caille}
\bse
\begin{eqnarray}
C^{3d}_{u}(\xv_\perp,z)&=&2T\int{d^2{k_\perp}d
k_z\over(2\pi)^3}{1-e^{i{\bf k}\cdot{\xv}}\over K  k_\perp^4 + B
k_z^2}
\equiv{T\over2\pi\sqrt{K  B}}\;g^{3d}_T\left({z\lambda\over x_\perp^2},
{x_\perp\over a}\right)\;\\
&=&{T\over2\pi\sqrt{K  B}}
\left[\ln\left({x_\perp\over a}\right)-
\frac{1}{2}{\text{Ei}}\left({-x_\perp^2\over 4\lambda|z|}\right)\right],
\ \ \ \ \ \ \ \ \ \\
&\approx&{T\over2\pi\sqrt{K  B}}\left\{\begin{array}{lr}
\ln\left({x_\perp\over a}\right),&x_\perp\gg\sqrt{\lambda|z|}\;,\\
\ln\left({4\lambda z\over a^2}\right),&x_\perp\ll\sqrt{\lambda|z|}\;,\\
\end{array}\right.
\label{Cuu3dT0}
\end{eqnarray}
\ese
where $\text{Ei}(x)$ is the exponential-integral function.  As
indicated in the last form, in the asymptotic limits of
$x_\perp\gg\sqrt{\lambda z}$ and $x_\perp\ll\sqrt{\lambda z}$ above 3d
correlation function reduces to a logarithmic growth with $x_\perp$
and $z$, respectively.

In 2d one instead finds\cite{TonerNelsonSm}
\bse
\begin{eqnarray}
C^{2d}_{u}(x,z)&=&2T\int{d{k_x}d
k_z\over(2\pi)^2}{1-e^{i{\kv}\cdot{\xv}}\over K  k_x^4 + B k_z^2}
\equiv {T\over2\pi\sqrt{K  B}}\;g^{2d}_T\left({z\lambda\over x^2},
{x\over a}\right)\;,\\
&=&{2T\over B}
\bigg[\left(\frac{|z|}{4\pi\lambda}\right)^{1/2}
  e^{-x^2/(4\lambda|z|)}
+ \frac{|x|}{4\lambda}\mbox{erf}
\big(\frac{|x|}{\sqrt{4\lambda|z|}}\big)\bigg]\ \ \ \ \ \ \ \\
&\approx&{2T\over B}\left\{\begin{array}{lr}
\left(\frac{|z|}{4\pi\lambda}\right)^{1/2},
&x\ll\sqrt{\lambda|z|}\;,\ \ \ \ \\
\frac{|x|}{4\lambda}, &x\gg\sqrt{\lambda|z|}\;,\ \ \ \ \\
\end{array}\right.
\label{Cuu2dT0}
\end{eqnarray}
\ese
where $\text{erf}(x)$ is the Error function. 

As a consequence of above divergent phonon fluctuations, the smectic
density wave order parameter \rf{rhoSm} vanishes in
thermodynamic limit
\bse
\begin{eqnarray}
\langle\rho(\xv)\rangle_0
&=&2|\rho_q|\langle 
\cos\big[\qv_0\cdot\xv - q u(\xv)\big]\rangle_0,\\
&=&2|\rho_q|e^{-\oh q_0^2\langle u^2\rangle_0}
\cos\big(\qv_0\cdot\xv),\\
&=&2\tilde\rho_q(L_\perp)\cos\big(\qv_0\cdot\xv),
\label{DeltaLOave2b}
\end{eqnarray}
\ese
with the thermally suppressed order parameter amplitude given by
\bse
\begin{eqnarray}
\hspace{-1cm}
\tilde\rho_{q}(L_\perp)&=&
|\rho_{q}|
\left\{\begin{array}{ll}
e^{-L_\perp/\xi_\perp},& d = 2,\\
\left(\frac{a}{L_\perp}\right)^{\eta/2},& d = 3,\\
\end{array}\right.\\
&\rightarrow &0,\ \ \mbox{for $L_\perp\rightarrow\infty$},
\label{DeltaR}
\end{eqnarray}
\ese 
where I used results for the phonon and phase fluctuations, \rf{uuT},
and defined the Caill\'e exponent
\begin{eqnarray}
  \eta&=&\frac{q_0^2 T}{8\pi\sqrt{B K}}.
\end{eqnarray}
Thus, in qualitative contrast to its mean-field cartoon, at long
scales (longer than $\xi_{\perp,z}$) the smectic state is actually
characterized by a {\em uniform} mass density.

Since the average density is actually uniform, a better
characterization of the smectic state is through the structure
function, $S(\qv)$, a Fourier transform of the density correlation
function, that in 3d is given by
\bse
\begin{eqnarray}
S(\qv)&=&\int d^3x\langle\delta\rho(\xv)\delta\rho(0)\rangle e^{-i\qv\cdot\xv},\\
&\approx&\oh\sum_{q_n}|\rho_{q_n}|^2\int_{\xv}
\langle e^{-i q_n (u(\xv) - u(0))}\rangle_0
  e^{-i(\qv - q_n\zh)\cdot \xv},\ \ \ \ \ \ \ \ \ \\
&\approx&\oh\sum_{n}\frac{|\rho_{q_n}|^2}{|q_z - n q_0|^{2-n^2\eta}}, 
\ \ \mbox{for $d=3$},
\end{eqnarray}
\label{Sq}
\ese
where I approximated phase and phonon fluctuations by Gaussian
statistics (in 3d valid up to weak logarithmic corrections\cite{GP}).
Thus as anticipated one finds that the logarithmically divergent 3d
phonon fluctuations lead to a structure function, with highly
anisotropic ($q_z\sim q_\perp^2/\lambda$) {\em quasi}-Bragg peaks
replacing the true ($\delta$-function) Bragg peaks characteristic of a
true long-range periodic order.\cite{deGennes,ChaikinLubensky,SqSmExp}

In two dimensions, smectic order is even more strongly suppressed by
thermal fluctuations. The linear growth of the 2d phonon fluctuations
leads to exponentially short-ranged correlations of the density,
expected to result in dislocation unbinding at any nonzero
temperature, thereby completely destroying smectic state in
2d.\cite{TonerNelsonSm}

\subsection{Nonlinear elasticity: beyond Gaussian fluctuations}
\label{elastRG}
\subsubsection{Perturbation theory}
As is clear from the analysis of the previous subsection, the
restoration of the translational symmetry (a vanishing
$\langle\rho_{q}(\xv)\rangle$, etc.) by thermal fluctuations is a
robust prediction of the quadratic theory, that cannot be overturned
by the left-out nonlinearities. However, as discussed in the
Introduction and illustrated instability of the smectic Gaussian fixed
point (in $d<3$) in Fig.\ref{criticalPhaseFixedPointFig}, the
asymptotic long-scale form of the correlation functions computed
within the harmonic approximation only extends out to the nonlinear
length scales $\xi_{\perp,z}^{NL}$.  On longer scales, the divergently
large smectic phonon fluctuations invalidate the neglect of phonon
nonlinearities,

\begin{eqnarray}
  \cH_{\text{nonlinear}}&=&-\oh B (\partial_z u)(\nabla u)^2 
  +\frac{1}{8} B (\nabla u)^4.\ \ \ \
\label{Hnonlin}
\end{eqnarray}
These will necessarily qualitatively modify predictions \rf{Cuu3dT0},
\rf{Cuu2dT0}, and \rf{Sq} on scales longer than the crossover scales
$\xi^{NL}_{\perp,z}$, that I compute next.
%
\begin{figure}[htbp]
\hspace{0in}\includegraphics*[width=0.3\textwidth]
{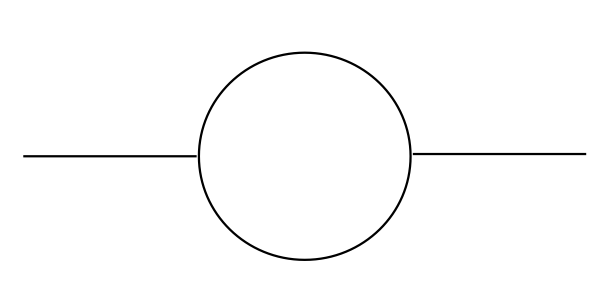}
\caption{Feynman graph that renormalizes the elastic moduli $K$, $B$
  of the smectic state.}
\label{fig:deltaBloop}
\end{figure}

To see this, one uses a perturbative expansion in the nonlinear
operators \rf{Hnonlin} to assess the size of their contribution to
e.g., the free energy. Following the standard field-theoretic RG
analysis, pioneered by Wilson and Fisher\cite{epsilonWilsonFisher}
(but here applied to the critical smectic phase, rather than to a
critical point) these can be accounted for as corrections to the
compressional $B$ and bend $K$ elastic moduli, with the leading
softening (negative) contribution to $\delta B$, summarized
graphically in Fig.\ref{fig:deltaBloop}, and given by
\bse
\begin{eqnarray}
\delta B&=&-\oh T B^2\int_{\bf q} q_\perp^4
G_{u}({\bf q})^2\;,\label{deltaBa}\\
&\approx&-\oh T B^2\int_{-\infty}^{\infty}{d q_z\over2\pi}
\int_{L^{-1}_\perp}{d^{d-1}q_\perp\over(2\pi)^{d-1}}
{q_\perp^4\over(K q_\perp^4+B q_z^2)^2}\;,\label{deltaBb}\nonumber\\
&\approx&-{1\over8} {C_{d-1}T\over 3-d}\;
\left({B\over K^3}\right)^{1/2} L_\perp^{3-d}B\;.\label{deltaBc}
\end{eqnarray}
\ese
In above, I used the smectic correlator, $G_{u}(\qv)$, focused on
$d\leq3$ (which allowed the dropping of the uv-cutoff ($\Lambda$)
dependent part that vanishes for $\Lambda\rightarrow\infty$), and
cutoff the divergent contribution of the long wavelength modes via the
infra-red cutoff $q_\perp>1/L_\perp$ by considering a system of a
finite extent $L_\perp$.  Similar analysis gives a {\em positive}
fluctuation correction to the bend modulus $K$, stiffening the
undulation mode.\cite{ffloLR}

Clearly the anharmonicity becomes important when the fluctuation
corrections to the elastic constants (e.g., $\delta B$ above) become
comparable to its bare microscopic value. As the analysis below
demonstrates, the divergence of this correction as
$L_\perp\rightarrow\infty$ signals the breakdown of the conventional
harmonic elastic theory on length scales longer than a crossover scale
\begin{eqnarray}
\xi^{NL}_\perp&\approx&
\left\{\begin{array}{ll}
\frac{1}{T}\left(\frac{K^3}{B}\right)^{1/2},& d = 2,\\
a e^{\frac{c}{T}\left(\frac{K^3}{B}\right)^{1/2}},& d = 3,
\end{array}\right.
\label{xiNL}
\end{eqnarray}
which I define here as the value of $L_\perp$ at which
$|\delta B(\xi^{NL}_\perp)|=B$. Within the approximation of the
smectic screening length $\lambda = a$, these nonlinear crossover
lengths reduce to the phonon disordering lengths \rf{xi_pxi_z},
\rf{xiperp}, defined by the Lindemann-like criterion.\cite{Lindemann}
Clearly, on scales longer than $\xi^{NL}_{\perp,z}$ (also a crossover
scale from Gaussian to smectic fixed point in
Fig.\ref{conventionalPhaseFixedPointFig}) the perturbative
contributions of nonlinearities dominate and therefore cannot be
neglected. Their contribution are thus expected to qualitatively
modify the harmonic predictions of the previous subsection.

%
%

\subsubsection{Renormalization group analysis in $d=3-\epsilon$ dimensions} 
\label{sec:RG}

To describe the physics beyond the crossover scales,
$\xi^{NL}_{\perp,z}$ -- i.e., to make sense of the infra-red divergent
perturbation theory found in Eq.\ref{deltaBc}\ -- requires a
renormalization group analysis.\cite{epsilonWilsonFisher, MEFrmp74, WilsonKogutPR} This
was first performed in the context of conventional 3d smectic liquid
crystals and Lifshitz points in a seminal work by Grinstein and
Pelcovits (GP)\cite{GP}. Below, I complement GP's treatment with
Wilson's momentum-shell renormalization group (RG) analysis, extending
it to an arbitrary dimension $d$, so as to connect to the behavior in
2d, that has an exact solution\cite{GW}.

To this end I integrate (perturbatively in $\cH_{\text{nonlinear}}$)
short-scale Goldstone modes in an infinitesimal cylindrical shell of
wavevectors, $\Lambda e^{-\delta\ell}<q_\perp<\Lambda$ and
$-\infty<q_z<\infty$ ($\delta\ell\ll 1$ is infinitesimal).  The
leading perturbative momentum-shell coarse-graining contributions come
from terms found in direct perturbation theory above, but with the
system size divergences controlled by the infinitesimal momentum
shell. The thermodynamic averages can then be equivalently carried out
with an effective coarse-grained Hamiltonian of the form \rf{Hu_el},
with momentum-dependent couplings obtained from integrating up their
infinitesimal corrections, obtained via momentum-shell RG. For smectic
moduli $B$ and $K$ this gives,
\begin{eqnarray}
\delta B&\approx&-\frac{1}{8} g B\delta\ell,\ \ \ 
\delta K\approx\frac{1}{16} g K\delta\ell,
\label{deltaBdeltaK}
\end{eqnarray}
where the dimensionless coupling is given by
\begin{eqnarray}
g&=&C_{d-1}\Lambda_\perp^{d-3}T\left({B\over K^3}\right)^{1/2}
\approx\frac{T}{2\pi}\left({B\over K^3}\right)^{1/2}\;,
\label{g}
\end{eqnarray}
and in the second form I approximated $g$ by its value in 3d.
Eqs.\rf{deltaBdeltaK} show that $B$ is softened and $K$ is stiffened
by the nonlinearities in the presence of thermal fluctuations, making
the system effectively more isotropic, as one may expect on general
physical grounds.

For convenience I then anisotropically rescale the lengths and the
remaining long wavelength part of the fields $u^<({\bf r})$ according
to $r_\perp=r_\perp'e^{\delta\ell}$, $z=z'e^{\omega\delta\ell}$ and
$u^<({\bf r})= e^{\phi\delta\ell}u'({\bf r'})$, so as to restore the
ultraviolet cutoff $\Lambda_\perp e^{-\delta\ell}$ back up to
$\Lambda_\perp$. The underlying rotational invariance ensures that the
nonlinear fluctuation corrections preserve the rotationally invariant
strain operator $\big(\partial_z u - \oh({\nabla}_\perp u)^2\big)$,
renormalizing it as a whole. It is therefore convenient (but not
necessary) to choose the dimensional rescaling that also preserves
this form. It is easy to see that this choice leads to
\begin{equation}
\phi=2-\omega\;.\label{chi_choice}
\end{equation}
The leading (one-loop) corrections to the effective coarse-grained and
rescaled free energy functional can then be summarized by differential
RG flows
\bse
\begin{eqnarray}
\frac{d B(\ell)}{d\ell}&=&(d+3-3\omega-{1\over8}g(\ell))B(\ell)
\;,\label{Bflow}\\
\frac{d K(\ell)}{d\ell}&=&(d-1-\omega+{1\over16}g(\ell))K(\ell)
\;.\label{Kflow}
\end{eqnarray}
\ese 
From these one readily obtains the flow of the dimensionless coupling
$g(\ell)$
\begin{eqnarray}
\frac{d g(\ell)}{d\ell}&=&(3-d)g-\frac{5}{32}g^2\;,\label{g_flow}
\end{eqnarray}
whose flow for $d<3$ away from the $g=0$ Gaussian fixed point encodes
the long-scale divergences found in the direct perturbation theory
above. This resembles the Wilson-Fisher flow at the critical point
just below $d = 4$\cite{epsilonWilsonFisher}, but, as discussed in the
Introduction requires no fine tuning. As summarized in
Fig.\ref{criticalPhaseFixedPointFig} for $d<3$ the flow terminates at
a nonzero fixed-point coupling $g_*=\frac{32}{5}\epsilon$ (with
$\epsilon\equiv 3-d$), that determines the nontrivial long-scale
behavior of the smectic critical phase (see below). As with treatments
of critical points\cite{epsilonWilsonFisher, ChaikinLubensky}, but
here extending over the whole smectic phase, the RG procedure is
quantitatively justified by the proximity to $d=3$, i.e., controlled
by the smallness of $\epsilon$.

One can now use a standard matching calculation to determine the
long-scale asymptotic form of the correlation functions on scales
beyond $\xi^{NL}_{\perp,z}$. Namely, applying above coarse-graining RG
analysis to a computation of correlation functions allows one to
relate its form at long length scales of interest (that, because of
infrared divergences is impossible to compute via a direct
perturbation theory in nonlinearities) to its counterpart at short
scales, evaluated with coarse-grained couplings, $B(\ell)$,
$K(\ell)$,\ldots. In contrast to the former, the latter is readily
computed via a perturbation theory, that, because of shortness of the
length scale is convergent. The result of this matching calculation to
lowest order gives correlation functions from an effective Gaussian
theory,
\begin{eqnarray}
G_{u}(\kv)
&\approx&\frac{T}{B(\kv) k_z^2 + K(\kv) k_\perp^4},
\label{Guu}
\end{eqnarray}
with moduli $B(\kv)$ and $K(\kv)$ that are singularly
wavevector-dependent, determined by the solutions $B(\ell)$ and
$K(\ell)$ of the RG flow equations \rf{Bflow} and \rf{Kflow}, with
initial conditions given by the microscopic values $B$ and $K$.

{\em 2d analysis:} In $d=2$, at long scales $g(\ell)$ flows to a
nontrivial infrared stable fixed point $g_*=32/5$, and the matching
analysis predicts correlation functions characterized by anisotropic
wavevector-dependent moduli
\bse
\begin{eqnarray}
K({\bf k})&=&K\left(k_\perp\xi^{NL}_\perp\right)^{-\eta_K}
f_K(k_z\xi^{NL}_{z}/(k_\perp\xi^{NL}_\perp)^\zeta)\;,\label{Kg}
\ \ \ \ \ \ \ \ \ \\
&\sim& k_\perp^{-\eta_K},\nonumber\\
B({\bf k})&=&B\left(k_\perp\xi^{NL}_\perp\right)^{\eta_B}
f_B(k_z\xi^{NL}_{z}/(k_\perp\xi^{NL}_\perp)^\zeta)\;,\label{Bg}\\
&\sim& k_\perp^{\eta_B}.\nonumber
\end{eqnarray}
\label{KgBg}
\ese
Thus, on scales longer than $\xi^{NL}_{\perp,z}$ these qualitatively
modify the real-space correlation function asymptotics of the harmonic
analysis in the previous subsection.  In Eqs.\rf{KgBg} the universal
anomalous exponents are given by
\bse
\begin{eqnarray}
\eta_B&=&{1\over8}g_*={4\over 5}\;\epsilon\;,\label{etaB2}\nonumber\\
&\approx&{4\over5}\;,\;\;\;\mbox{for}\; d=2\;,\label{etaB3}\\
\eta_K&=&{1\over16}g_*={2\over 5}\;\epsilon\;,\label{etaK2}\nonumber\\
&\approx&{2\over5}\;,\;\;\;\mbox{for}\; d=2\;,\label{etaK3}
\end{eqnarray}
\ese 
determining the $z-\xv_\perp$ anisotropy exponent via \rf{Guu} to be
\bse
\begin{eqnarray}
\zeta&\equiv& 2-(\eta_B+\eta_K)/2\;,\\
&=&\frac{7}{5},\;\mbox{for}\; d=2\;,\label{zeta3}
\end{eqnarray}
\label{zeta}
\ese
as expected reduced by thermal fluctuations down from its harmonic
value of $2$. The $\kv_\perp-k_z$ dependence of $B(\kv),K(\kv)$ is
determined by universal scaling functions, $f_B(x),f_K(x)$ that I will
not compute here. The underlying rotational invariance gives an {\em
  exact} relation between the two anomalous $\eta_{B,K}$ exponents
\bse
\begin{eqnarray}
3-d &=& {\eta_B\over 2} + {3\over  2}\eta_K\;,
\label{WI}\\
1  &=& {\eta_B\over 2} + {3\over 2}\eta_K\;,\ \ \mbox{for
  $d=2$},
\label{WI2d}
\end{eqnarray}
\ese
which is obviously satisfied by the anomalous exponents,
Eqs.\rf{etaK2},\rf{etaB2}, computed here to first order in
$\epsilon=3-d$.

Thus, as advertised, I find that at nonzero temperature, a 2d smectic
state is highly nontrivial and qualitatively distinct from its
mean-field perfectly periodic form. In addition to a vanishing density
modulation and associated fluctuation-restored translational symmetry,
it is characterized by universal nonlocal length-scale dependent
moduli, \rfs{KgBg}. Consequently its Goldstone mode theory and the
associated correlations are not describable by a local harmonic field
theory, that is an analytic expansion in local field operators.
Instead, in 2d, on length scales beyond $\xi^{NL}_{\perp,z}$ (but
shorter than the expected dislocation unbinding length $\xi_{disl}$)
thermal fluctuations and correlations of this smectic {\em critical}
phase are controlled by a nontrivial fixed point, illustrated in
Fig. \ref{criticalPhaseFixedPointFig}, characterized by {\em universal}
anomalous exponents $\eta_{K,B}$ and scaling functions $f_{B,K}(x)$,
defined above.

Above I obtained this nontrivial structure from an RG analysis and
estimated these exponents within a controlled but approximate
$\epsilon$-expansion. Remarkably, in 2d an exact solution of this
problem was discovered by Golubovic and Wang\cite{GW}. It predicts
critical phenomenology in a qualitatively agreement with the RG
predictions above, and gives exact exponents
%
\begin{eqnarray}
\eta_B^{2d}&=&1/2,\ \ \eta_K^{2d}=1/2,\ \ \zeta^{2d}=3/2,
\label{etaGW2d}
\end{eqnarray}
%
derived by maping onto the 1+1d KPZ equation.\cite{KPZ}

{\em 3d analysis:} In $d=3$, the nonlinear coupling $g(\ell)$ is
marginally irrelevant, flowing to $0$ at long scales. Despite this, as
discovered by Larkin and Khmel'nitskii and by Michael Fisher, et al.,
in the context of the Ising model\cite{LarkinKhm69, MEFmarginal72,
  MEFrmp74}, the marginal flow to the Gaussian fixed point is
sufficiently slow (logarithmic in lengths) that (as usual at a
marginal dimension\cite{ChaikinLubensky}) its power-law in $\ell$
dependence leads to a universal, asymptotically {\em exact}
logarithmic wavevector dependence found in a smectic by Grinstein and
Pelcovits\cite{GP},
\bse
\begin{eqnarray}
  K({\kv_\perp,k_z=0})&\sim&K|1+
\frac{5g}{64\pi}\ln(1/k_\perp a)|^{2/5}\;,\label{K3d}\\
  B({\kv_\perp=0,k_z})&\sim&B|1+\frac{5g}{128\pi}
\ln(\lambda/k_za^2)|^{-4/5}.\ \ \ \ \ \ \ \ \ \ \label{B3d}
\end{eqnarray}
\label{KB3d}
\ese
This translates into the smectic order parameter correlations given by
\begin{eqnarray}
\langle\rho^*_{q}(\xv)\rho_{q}(0)\rangle&\sim&e^{-c_1(\ln z)^{6/5}}\cos(q_0 z),
\label{SrGP}
\end{eqnarray}
with $2/5, 4/5$, and $6/5$ {\em exact} universal exponents and $c_1$
a nonuniversal constant\cite{GP}.  Although these 3d anomalous effects
are less dramatic and likely to be difficult to observe in practice,
theoretically they are quite significant as they represent a
qualitative breakdown of the mean-field and harmonic descriptions,
that respectively ignore interactions and thermal fluctuations.


I conclude this section by noting that all of the above analysis is
predicated the validity of the purely elastic model, \rfs{Hu_el}, that
neglects topological defects, such as dislocations. If these unbind
(as they undoubtedly do in 2d at any nonzero
temperature\cite{TonerNelsonSm}), then above prediction only hold on
scales shorter than the separation $\xi_{disl}$ between these defects.

\section{Biaxially periodic critical states: spontaneous line-crystals}

Another class of critical states are those partially crystalized along
{\em two} of the three axes. They can be dubbed as ``{\em spontaneous}
line-crystals'' as they are a periodic 2d crystalline array of 1d
liquids, oriented along a {\em spontaneously} chosen axis, illustrated
in Fig.\ref{criticalPhasesFig}(c), and in this sense dual to a
uniaxially periodic critical state -- a 1d periodic array of 2d liquids
-- the $m=d-1$-Lifshitz model\cite{mLifshitz}, discussed in
Sec.\ref{uniaxialSec}.  As I discuss below, its key feature is
rotational symmetry-enforced vanishing of the tilt modulus -- the
$m=1$-Lifshitz model\cite{mLifshitz}, that leads to its enhanced
critical fluctuations.

\subsection{Columnar liquid crystal}

One example of a spontaneous line-crystal is a ubiquitous ``columnar''
liquid crystal. It emerges from a nematic fluid composed of a high
aspect ratio disk-shaped constituents, illustrated in
Fig. \ref{criticalPhasesFig}(c).\cite{deGennes} In this phase
disk-shaped molecules stack into one-dimensional fluid columns along a
spontaneous nematic axis, that freeze into a 2d crystal (typically
triangular lattice), retaining fluid order along the columns. Such
discotic liquid crystal, forming a 2d crystal of 1d fluid columns
compliments the smectic state of 1d periodic array of 2d fluids,
discussed in earlier sections.

Simple analysis similar to that of a smectic, shows that for $d < 5/2$
the columnar liquid crystal ($m=1$-Lifshitz model\cite{mLifshitz}) is
a critical phase, though given this dimensional constraint not a very
practical one for experimental realization.  However, although I do
not pursue the subject here, I note that far stronger distortion
effects of quenched disorder (e.g., columnar state in
aerogel\cite{RTaerogelPRL, RTaerogelPRB, BRTCaerogelScience}) lead to
a critical columnar glass for $d < 7/2$, characterized by an infrared
attractive fixed point controlled by disorder.\cite{SRTcolumnar}

The columnar state is characterized by a two-component phonon field
$\uv = (u_x,u_y)$, that are two Goldstone modes associated with
translational symmetry breaking in the plane transverse to the
columns. Analogously to a smectic, the Goldstone modes associated with
the rotational symmetry are gapped out by the emergent Higgs
mechanism.\cite{cholestericLR}

Although one can derive the Goldstone mode elasticity for the columnar
state by following the approach analogous to that of a smectic, here I
will simply write it down based on symmetry and experience with the
smectic. To this end I note that the columnar state exhibits two types
of rigidities, the in-plane (transverse to the columnar axis $\zh$)
crystalline elasticity and bending elasticity of the columns. The
former is characterized by shear and bulk elasticity of a 2d
crystal. The latter is captured by the higher-derivative curvature
(rather than tension) filament elasticity. Together these give the
energy density
\begin{eqnarray}
\cH_{col} = \frac{1}{2}\kappa(\partial_z^2{\bf u})^2 +
\frac{\lambda}{2}\,u_{\alpha\alpha}^2 + \mu\, u_{\alpha\beta}^2,
\label{Hcol}
\end{eqnarray}
where $\kappa$ is the curvature modulus, $\mu$ and $\lambda$ are
Lam\'e elastic moduli of 2d triangular lattice\cite{ChaikinLubensky},
and (with the summation convention over repeated indices) the strain
tensor is given by
\begin{eqnarray}
u_{\alpha\beta} = \frac{1}{2}\,\Big(\partial_\alpha u_\beta 
+ \partial_\beta u_\alpha - \partial_\gamma
u_\alpha\partial_\gamma u_\beta\Big)\approx 
 \frac{1}{2}\,\Big(\partial_\alpha u_\beta + \partial_\beta u_\alpha - \partial_z
  u_\alpha\partial_z u_\beta\Big).
\label{u_nonlinear} 
\end{eqnarray}
The two phonons, $\uv(x,y,z) = (u_x,u_y)$ are fields in a
three-dimensional space, and in the second approximate form above I
only kept the most important strain nonlinearities.

As in the discussion of the smectic elasticity, the columnar nonlinear
elasticity \rf{Hcol} is strictly constrained by symmetry. The linear
$z$ derivative in the harmonic terms is forbidden by the rotational
invariance about the $x$ axes (broken spontaneously), at infinitesimal
level corresponding to $\uv\rightarrow \uv + \theta z{\bf \yh}$. This
``softness'' with respect to $\partial_z\uv$ modes then requires one
to keep corresponding nonlinearities in the in-plane strain tensor,
$u_{\alpha\beta}$ above, the form of which is dictated by the in-plane
rotational invariance.

As in a smectic, here too the analysis of quadratic phonon
fluctuations leads to power-law divergence for $d < 5/2$, requiring
inclusion of elastic nonlinearities in \rf{Hcol} via
\rf{u_nonlinear}. Account of these via an RG analysis very similar to
that of the smectic in the previous section, leads to a nontrivial
infrared stable fixed point controlled by $\epsilon = 5/2-d$, that
characterizes the universal properties of the resulting columnar {\em
  critical phase} (as in Fig.\ref{criticalPhaseFixedPointFig}).  For
technical details, I refer the reader to the original
literature.\cite{SRTcolumnar}

\subsection{Spontaneous vortex lattice in magnetic superconductors}

Another putative realization of a biaxially periodic (line-crystal)
critical state is a ``spontaneous vortex lattice''' in a ferromagnetic
supercondutor (FS).  Rare-earth borocarbide materials are believed to
be examples of the latter, exhibiting a rich phase diagram that
includes superconductivity, antiferromagnetism, ferromagnetism and
spiral magnetic order.\cite{PhysicsToday,experiments,Varma} In
particular, there is now ample experimental evidence that, at low
temperatures, superconductivity and ferromagnetism competitively
coexist in ErNi$_2$B$_2$C compounds, and perhaps in more recently
discovered high temperature superconductor Sr$_2$Y
Ru$_{1-x}$Cu$_x$O$_6$.

For sufficiently strong ferromagnetism, such FS's have been
predicted\cite{Varma} to exhibit a {\em spontaneous} vortex (SV) state
driven by the spontaneous magnetization, rather than by an external
magnetic field ${\bf H}$.  It is clear based purely on symmetry
arguments that for ${\bf H}=0$ and in the absence of atomic crystal
anisotropy, the elastic properties of the resulting SV solid differ
dramatically and {\em qualitatively} from those of a conventional
Abrikosov lattice. The key underlying difference is the aforementioned
{\em vanishing} of the tilt modulus \rf{Hcol}, which is guaranteed by
the underlying rotational invariance. Although this invariance is
broken by the magnetization, ${\bf M}$, the tilt modulus remains zero
because this breaking is {\em spontaneous} (neglecting the
ever-present crystal field anisotropy).  This contrasts strongly with
a conventional vortex solid, where the rotational symmetry is {\em
  explicitly} broken by the {\em applied} field ${\bf H}$ and lattice
anisotropy.

Consequently, the elasticity of the SV crystal is identical to that of
the columnar phase, described by\rf{Hcol} together with
\rf{u_nonlinear}, neglecting crystalline anisotropy.  In the presence
of purely thermal fluctuations, elastic nonlinearities remain
irrelevant for such 3d SV state, since $d=3 > 5/2$. However, as
discovered and explored in
Refs. \onlinecite{magneticSCprl,magneticSCprb}, in the presence of
ever-present quenched disorder, SV is a disorder- (rather than
temperature-) controlled critical glass state, treatable via RG
perturbative in $\epsilon = 7/2-d$, with many interesting properties,
e.g., the unusually singular $B(H)$ relation.

\section{Polymerized membrane}

A fluctuating, {\em tensionless} and therefore curvature-(as opposed
to tension-) controlled membrane is a strongly fluctuating
system. Focus on such membranes was originally stimulated in part by
ubiquitous biophysical realizations (e.g., liposomes, cellular
membranes, cytoskeleton of a red blood cell, etc)\cite{JWSmembranes},
but more recently found an ideal realization in fluctuating graphene
sheets.\cite{graphene} The latter, polymerized (solid) membrane,
characterized by a nonzero in-plane shear rigidity, is particularly
fascinating case of a critical phase for its internal dimension
$D < 4$.

\begin{figure}[htbp]
  \hspace{0in}\includegraphics*[width=0.5\textwidth]
  {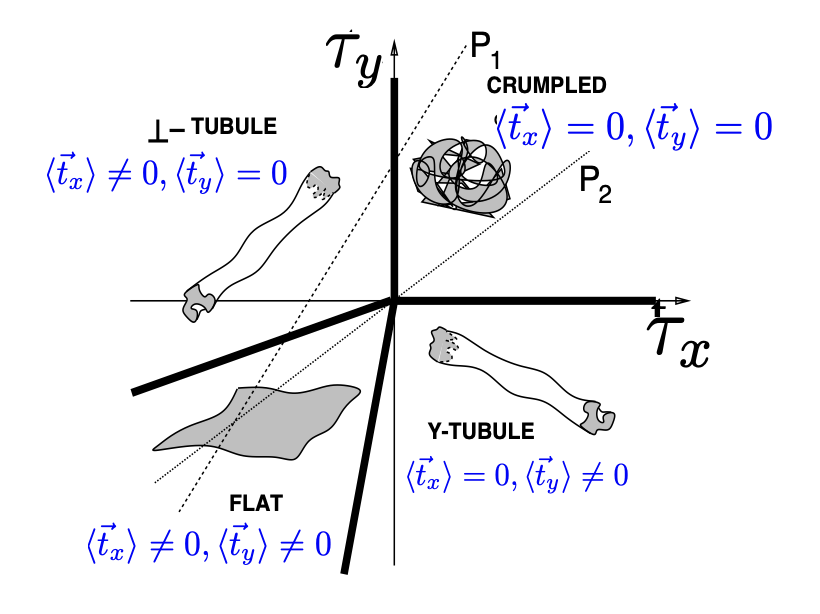}
  \caption{Illustration of a phase diagram of a polymerized membrane,
    controlled by temperature and in-plane anisotropy encoded in
    reduced temperatures $\tau_x$ and $\tau_y$. An isotropic membrane
    is described by $\tau_x = \tau_y\equiv\tau$, exhibiting
    temperature $\tau-$ driven transition along path $P_2$ from a
    crumpled phase (with vanishing tangent $\vec t_{x,y} =0$ order
    parameters) to a ``flat'' phase a phase with
    $\vec t_{x,y} \neq 0$.  Flat phase spontaneously breaks the
    rotational $O(3)$ symmetry of the embedding space and is a
    critical phase that exhibits rich universal power-law
    phenomenology, controlled by a nontrivial infrared stable fixed
    point.  A polymerized membrane with in-plane anisotropy is
    predicted to undergo a two-stage transition along $P_1$ from the
    crumpled-to-tubule phase (latter characterized by only one of the
    two tangent order parameters nonzero), followed by tubule-to-flat
    phase transition\cite{RTtubulePRL, RTtubulePRE}, later observed in
    simulations.\cite{tubuleSimulationsBowick}}
\label{membranePhaseDiagramFig}
\end{figure}

As was first argued by Nelson and Peliti\cite{NelsonPeliti}, in
addition to the rotationally invariant ``crumpled'' phase of linear
polymers and liquid membranes, polymerized membranes exhibit a finite
temperature ``flat'' phase. As I will discuss below, this too is a
{\em critical} phase characterized by universal power-law correlations
(e.g., membrane roughness) and anomalous elasticity, with the state's
very existence in two-dimensional membranes being a beautiful
illustration of a phenomena of order-from-disorder.
\begin{figure}[htbp]
  \hspace{-3.5in}\includegraphics*[width=0.3\textwidth]
  {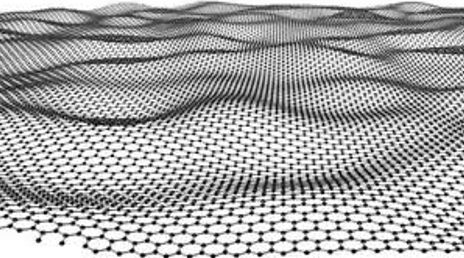}
  \put(35,-10){
    \includegraphics[width=0.5\textwidth]{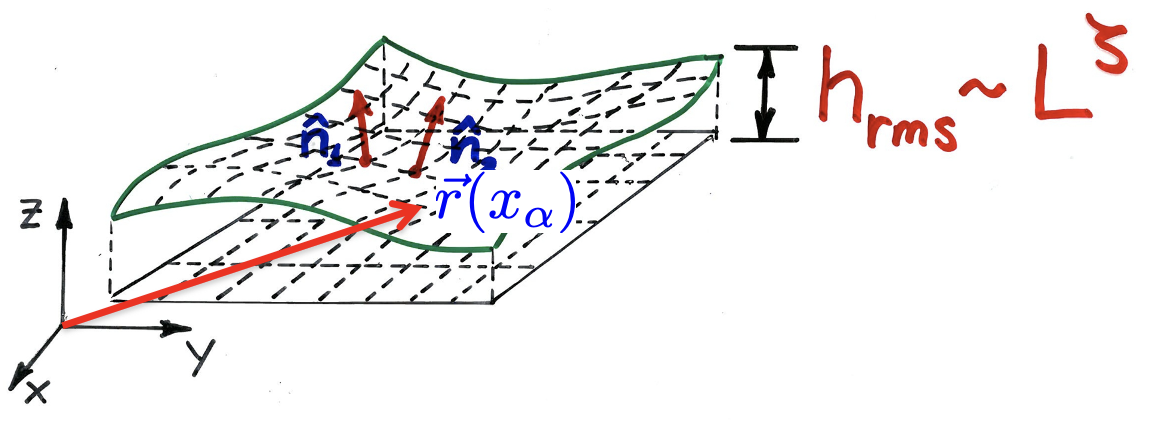}}
  \caption{Illustration of (a) a graphene sheet as a modern ideal
    realization of a thermally fluctuating polymerized membrane,
    predicted to exhibit rich anomalous elasticity as a critical
    phase, (b) flat phase of a polymerized membrane defining its
    normals, Monge gauge embedding and the definition of the roughness
    exponent, $\zeta$. }
\label{membraneFlatFig}
\end{figure}
I note in passing that, as illustrated in
Fig.\ref{membranePhaseDiagramFig}, a non-self-avoiding -- ``phantom''
anisotropic polymerized membrane was
predicted\cite{RTtubulePRL,RTtubulePRE} and observed in
simulations\cite{tubuleSimulationsBowick} to exhibit an intermediate
tubule phase, crumpled along one dimension and extended along the
other, that, despite its one-dimensionality, is stable to thermal
fluctuations.

A detailed derivation of the flat-phase Goldstone-modes elasticity is
available\cite{NelsonPeliti, JWSmembranes} starting from the Landau
theory of the crumpled phase in terms of the tangent vectors
$\grad\rvv(\xv)$, where $\rvv(\xv)$ is the embedding of the
$D$-dimensional membrane (parameterized by $\xv$) in the
$d$-dimensional space, $\rvv$ (with $D=2$ and $d=3$ the physical
case). The flat phase parameterized by
\begin{equation} 
  \rvv(\xv) =(t_0\xv
  + \uv(\xv), \hv(\xv)),
  \label{Monge}
\end{equation}
in the Monge gauge, spontaneously breaks the $O(d)$ symmetry of the
crumpled phase down to $O(D)$, where $t_0$ scale factor is the
effective order parameter of the flat phase, $\uv(\xv)$ is the
$D$-component in-plane phonon, and $\hv(\xv)$ is the ``height
function'' describing the membrane's transverse undulation into the
embedding space; we have implicitly generalized to arbitrary
co-dimension $d_c = d-D$.

The resulting energy functional is a sum of the bending and the
in-plane elastic contributions:
\begin{equation}
H_{flat}[\vec{h},{\bf u}]=\int d^Dx\left[ {\kappa \over 2} (\nabla^2\vec{h})^2
    + \mu u_{\alpha\beta}^2 +{\lambda \over 2} u_{\alpha\alpha}^2\right]\;,
\label{Fflat}
\end{equation}
where the nonlinear strain tensor is
\begin{eqnarray}
u_{\alpha \beta}&=&\oh(\partial_\alpha\rvv\cdot\partial_\beta\rvv -\delta_{\alpha\beta})
\approx {1\over 2} ( {\partial _{\alpha} u_{\beta}} +
{\partial _{\beta} u_{\alpha} } +
{\partial _{\alpha} {\vec h}}\cdot{\partial_{\beta} {\vec h}}),
\end{eqnarray}
defined as the deviation of the embedding-induced metric
$g_{\alpha\beta}$ from the flat metric, and in the second form I
neglected elastic nonlinearities that are subdominant at long
scales. A crucial feature of $H_{flat}[\vec{h},{\bf u}]$ \rf{Fflat}
(as with other critical phases) is the higher-order curvature
elasticity of $\hv$, with the surface tension modulus for
$(\grad\hv)^2$ vanishing exactly, enforced by the underlying
rotational invariance (of the embedding space) and tension-free
membrane.

One can also integrate out the noncritical in-plane phonon field
$\uv$, that appears only harmonically, to obtain a convenient
equivalent form, purely in terms of $\hv$ \cite{NelsonPeliti,
  AronovitzLubensky, LRphdThesis,LRmembraneAOP}
\begin{equation}
H_{flat}[\vec{h}]= \int d^Dx  [{\kappa \over 2} (\nabla^2\vec{h})^2
+{1 \over {4 d_c} }(\partial_\alpha\vec{h}\cdot\partial_\beta\vec{h}) 
R_{\alpha \beta, \gamma \delta}
(\partial_\gamma\vec{h}\cdot\partial_\delta\vec{h})]
\end{equation}
where for convenience, I rescaled Lam\'e coefficients so that the
quartic coupling is of order $1/d_c$.  The four-point coupling
fourth-rank tensor is given by
\begin{equation} 
R_{\alpha \beta, \gamma \delta}={K-2\mu\over2(D-1)} P^T_{\alpha \beta} 
P^T_{\gamma \delta}+{\mu\over2}\left(P^T_{\alpha \gamma} P^T_{\beta \delta } 
+ P^T_{\alpha \delta} P^T_{\beta \gamma}\right)\;,
\label{Rdefine}
\end{equation}
where
$P^T_{\alpha \beta}=\delta_{\alpha \beta} - q_{\alpha}q_{\beta}/q^2$
is a transverse (to ${\bf q}$) projection operator. The convenience of
this decomposition is that
$K= 2\mu (2 \mu + D \lambda)/(2 \mu + \lambda)$ and $\mu$ moduli
renormalize independently and
multiplicatively.\cite{LeDoussalRadzihovskyPRL, LRmembraneAOP}

I note that this last $h$-only elastic form indeed reflects the
general discussion of critical phases in the Introduction, namely that
they are described by an energy functional (of $\phi^4$ form with
$\vec\phi_\alpha\equiv\partial_\alpha\hv$) enforced to be critical
(i.e., missing the $\phi^2$ term) by the underlying rotational
symmetry, as illustrated in Fig.\ref{criticalPhaseFixedPointFig}.

Applying Wilson-Fisher momentum-shell RG analysis to membrane's
elastic
nonlinearities\cite{NelsonPeliti,AronovitzLubensky,DG,LeDoussalRadzihovskyPRL,
  LRmembraneAOP} one sees that for $D<4$ (and arbitrary embedding
dimension $d$) the membrane exhibits height undulations that diverge
with its extent $L$. As for smectics and other critical phases
discussed earlier, one can make sense of the associated divergent
perturbation theory in elastic nonlinearities by performing an RG
analysis controlled by $\epsilon = 4-D$\cite{AronovitzLubensky} or
equivalently $1/d_c$ (large embedding dimension, that is the analog of
$1/N$ large $N$ expansion of the $O(N)$ model's Wilson-Fisher critical
point)\cite{DG}.  The RG approach for arbitrary $d$ gives flows of the
dimensionless nonlinear coupling constants,
$\hat\mu(\ell) = \mu/\kappa^2, \hat\lambda(\ell) = \lambda/\kappa^2$,
after integrating out modes in a momentum shell
$\Lambda e^{-\delta\ell}< q <\Lambda$ (as for a smectic in previous
section), and rescaling for convenience.
%
%

The dimensionless couplings flow to a tuning-free infrared attractive
Aronovitz-Lubensky (AL) fixed
point,\cite{NelsonPeliti,AronovitzLubensky,DG,LeDoussalRadzihovskyPRL},
with two other (in addition to the Gaussian) fixed points that
describe a marginally mechanically unstable membrane (of physical
relevance to nematic elastomers of the next section).  The resulting
globally infrared stable AL fixed point controls the properties of the
highly nontrivial ``flat'' phase, that is critical and power-law
rough. A complementary approach of Le Doussal and
Radzihovsky\cite{LeDoussalRadzihovskyPRL} is via a solution of
self-consistent integral equations (SCSA) that build on the large
embedding $1/d_c$ expansion, with virtue of being exact in three
limits: (i) $D \rightarrow 4$ ($\epsilon$ expansion) for arbitrary
$d_c$, (ii) $d_c\rightarrow\infty$ ($1/d_c$ expansion) and (iii)
$d_c=0$ for arbitrary $D$, thereby providing best quantitative
predictions, as judged by numerics and experiments.

The ``flat'' critical phase is characterized by {\em universal
  anomalous} elasticity with length scale-dependent elastic moduli
\begin{eqnarray}
\kappa(q) \sim q^{-\eta},\ \ \mu(q)  \sim  \lambda (q) \sim q^{\eta_u},
\end{eqnarray}
where for physical membranes $\eta\approx 0.82$ and the underlying
embedding-space rotational invariance imposes exact relations
\cite{NelsonPeliti,AronovitzLubensky,DG,LeDoussalRadzihovskyPRL},
\begin{eqnarray}
\eta_u = 4-D - 2\eta_\kappa,\ \ \ \zeta = (4-D -\eta)/2.
\end{eqnarray}
These indicate that at long scales, $q\rightarrow 0$, thermal
fluctuations stiffen the effective bending modulus $\kappa(q)$ and
soften membrane's in-plane moduli, $\mu(q), \lambda(q)$. 

Based on these, the SCSA\cite{LeDoussalRadzihovskyPRL} predicts a {\em
  universal} and {\em negative} Poisson ratio,
\begin{equation}
\mathop{{\rm Lim}}\limits_{q\rightarrow 0}\;\;
\sigma\equiv{\lambda(q)\over
2\mu(q)+(D-1)\lambda(q)} =-{1\over 3}\;,\ \text{for}\ D=2,
\label{PoissonRatio}
\end{equation}
that measures the ratio of in-plane compression of a membrane along an
axis transverse to its strained direction. Its negative value
indicates that such membrane actually expands transversely as it is
being stretched. This approximate value of $-1/3$ compares extremely
well with largest simulations\cite{zetaBowick} and higher order
computations. This amazing fluctuation-driven anomalous elasticity
phenomenology, uncovered using systematic Wilson-Fisher RG, can be
understood qualitatively by playing around with a roughened piece of
paper.

Finally I conclude by emphasizing that the universal anomalous
elasticity is at the heart of stability of the 2D ``flat'' phase. To
see this, observe that in contrast to the unstable Gaussian fixed
point (describing fluctuations of a microscopic membrane), the
``flat'' {\em critical} phase at the AL fixed point, with a
renormalized, momentum-dependent $\kappa(q)$, is characterized by
root-mean-squared height undulations
\begin{eqnarray}
  h_{rms} & = & \sqrt{\langle h^2(\xv)\rangle}
=\left[\int_{L^{-1}}\frac{d^Dq}{(2\pi)^{d}}\frac{T}{\kappa(q) q^4}\right]^{1/2}
                \sim L^\zeta,
                \label{hrms}
\end{eqnarray}
with the roughness exponent $\zeta < 1$. Its best estimate for
$D=2,d=3$ \cite{LeDoussalRadzihovskyPRL} is $$\zeta = 0.59,$$ and has
also been computed within $\epsilon$ and $1/d_c$
expansions\cite{AronovitzLubensky,DG,LeDoussalRadzihovskyPRL,epsHigherOrder}.
I note that despite unbounded power-law height undulations \rf{hrms},
the {\em rotational} $O(d)$ symmetry of the embedding space remains
broken since $L^\zeta \ll L\rightarrow\infty$, despite
two-dimensionality ($D = 2$) of the polymerized membrane field theory.
It thus circumvents a naive application of the
Hohenberg-Mermin-Wagner-Coleman theorem\cite{Hohenberg, MerminWagner,
  Coleman}. The rigorous form of the latter does not a priori apply to
the higher-derivative polymerized membrane field theory, and thus does
not forbid above findings, e.g., stability of the 2D flat phase. More
physically, as can be seen from \rf{hrms}, theorem's common heuristic
application fails precisely because the state is controlled by a
non-Gaussian fixed point with anomalous, momentum-dependent elastic
moduli, specifically with $\eta > 0$.  As a result
$h_{rms}(L)/L \sim L^{-\eta/2}\rightarrow 0$ indicates that the
``flat'' phase is stabilized (for nonzero positive $\eta > 0$
generated by thermal fluctuations) by the very fluctuations that
attempt to destabilize it, a phenomenon known as order-from-disorder.

Finally, I note that random heterogeneity also ubiquitously appears in
most membrane realizations (see e.g., work on
graphene\cite{grapheneMceuen}) and leads to a disorder-driven critical
membrane phase, studied extensively starting with the work of Nelson
and Radzihovsky\cite{RadzihovskyNelsonEuro, RadzihovskyNelsonPRA},
with many that followed\cite{MorseLubensky, LRflatglassPRB,
  LRmembraneAOP}.

\section{Nematic elastomer}

\begin{figure}[htbp]
\hspace{0in}\includegraphics*[width=0.5\textwidth]
{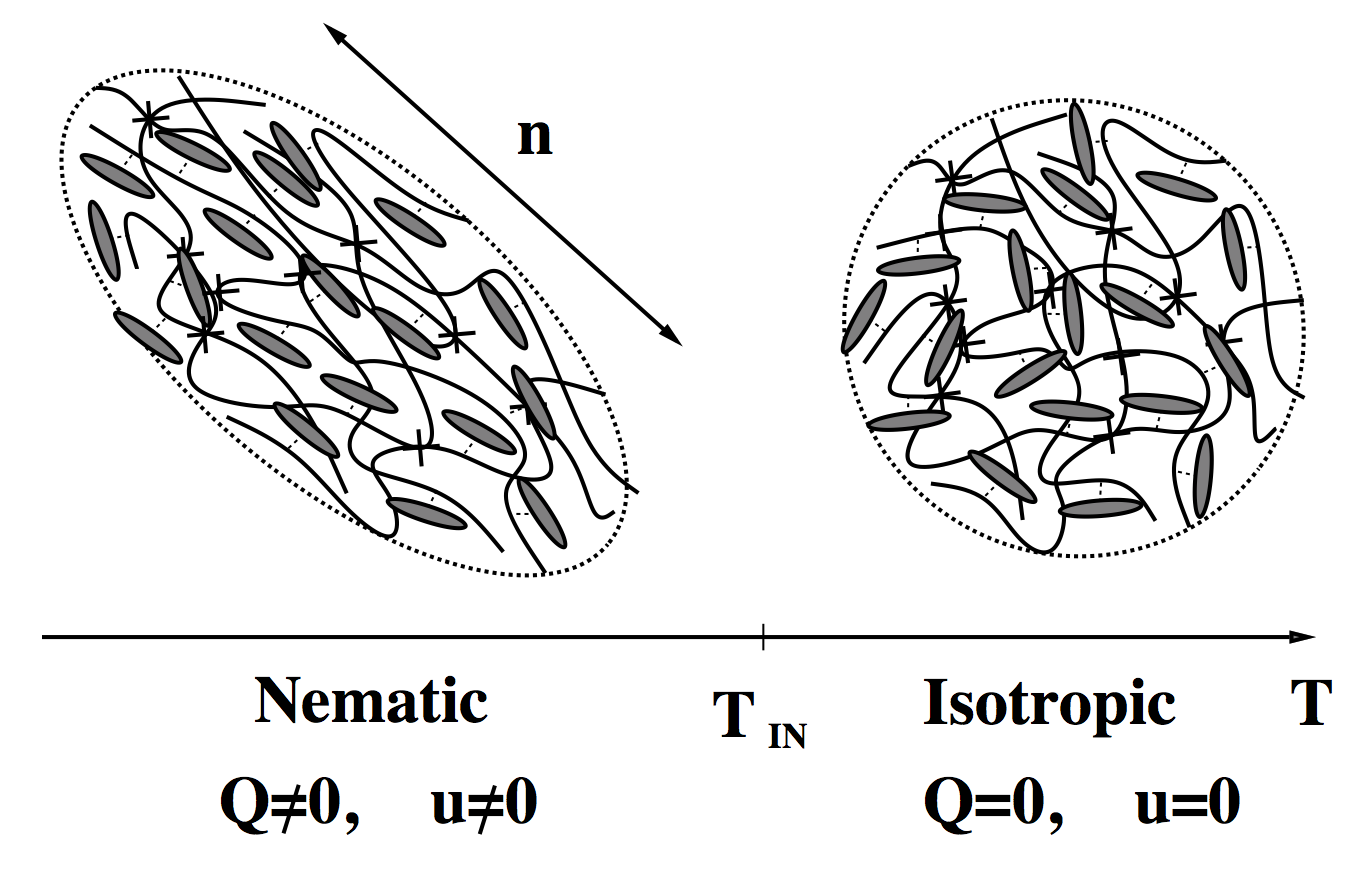}
\caption{Spontaneous uniaxial distortion of nematic elastomer driven
  through the isotropic-nematic transition.}
\label{elastomerINfig}
\end{figure}

A final example of a critical phase that I will briefly discuss is a
nematic elastomer, namely rubber -- a randomly crosslinked polymer
network composed of mesogenic groups, illustrated in
Fig.\ref{elastomerINfig}. It exhibits a spontaneous transition to a
nematic phase, thereby driving an accompanying spontaneous uniaxial
distortion of the elastic matrix, illustrated in
Fig.\ref{elastomerINfig}.\cite{bookWarner}

\begin{figure}[htbp]
\hspace{0in}\includegraphics*[width=0.7\textwidth]
{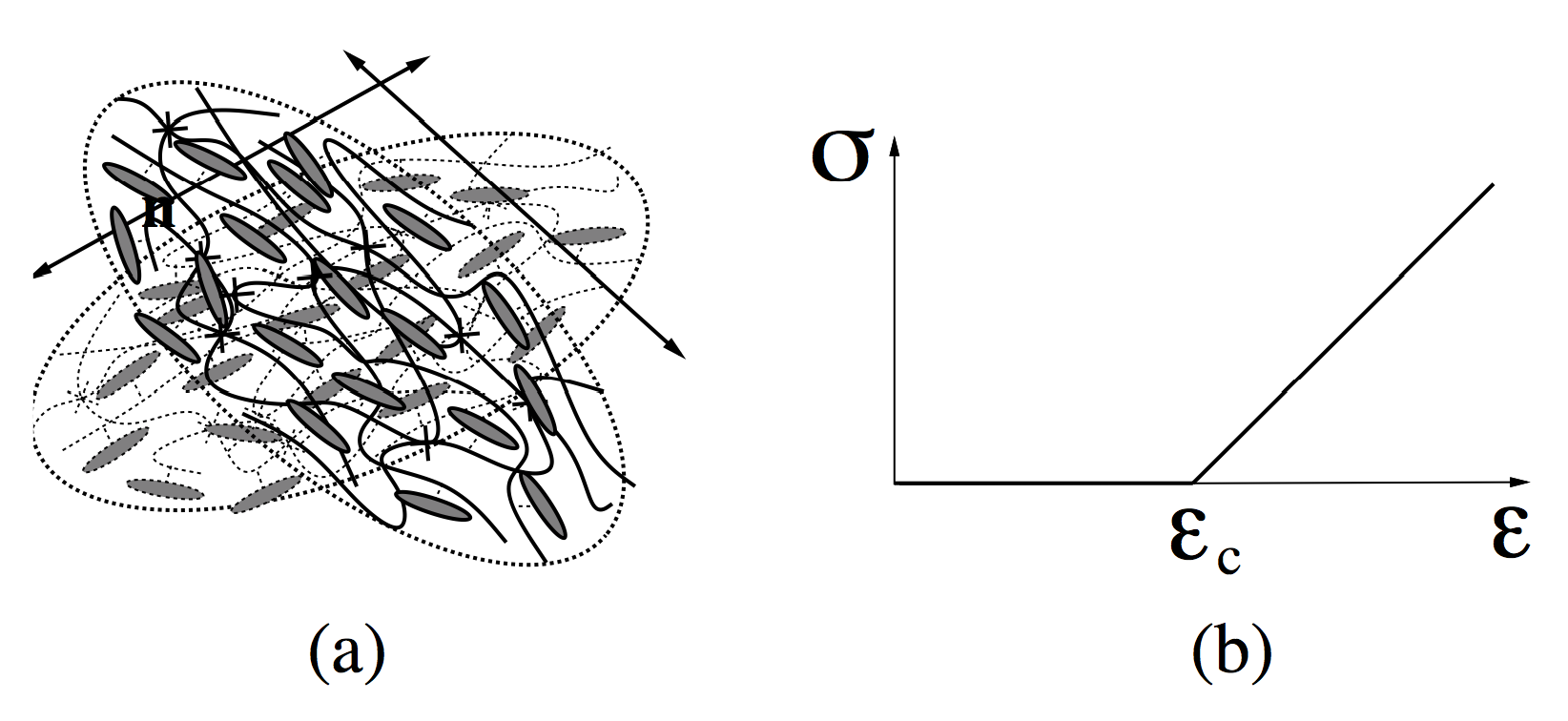}
\caption{(a) Simultaneous reorientation of the nematic director and of
  the uniaxial distortion is a low-energy nemato-elastic Goldstone
  mode of an ideal elastomer, that is responsible for its softness and
  (b) its resulting idealized flat (vanishing stress) stress-strain
  curve for a range of strains $0<\varepsilon <
  \varepsilon_c$. \cite{LMRX}}
\label{elastomerGoldstoneFig}
\end{figure}

Even in the absence of fluctuations, bulk nematic elastomers were
predicted\cite{GL} and later observed to display an array of
fascinating phenomena\cite{bookWarner,reviewNE}, the most striking of
which is the vanishing of stress for a range of strain, applied
transversely to the spontaneous nematic axis. This striking softness
is generic, stemming from the spontaneous orientational symmetry
breaking by the nematic state, accompanied by the nemato-elastic
Goldstone mode, that leads to the observed soft distortion and
strain-induced director reorientation\cite{GL,LMRX}, illustrated in
Fig.\ref{elastomerGoldstoneFig}. This unique elastic phenomenon is
captured even at the harmonic approximation of the fully nonlinear
uniaxial Hamiltonian in Eq.\rf{Hne},
\begin{eqnarray}
\hspace{-20mm}
\cH_{elast} &=&  \frac{1}{2} B_z \,w_{zz}^2 
+ \lambda_{z\perp}\, w_{zz}\, w_{\alpha\alpha} 
+ \frac{1}{2}\, \lambda\, w_{\alpha\alpha}^2  +  \mu\, w_{\alpha\beta}^2 
+ \frac{K}{2}\,(\nabla_{\perp}^2 u_z)^2, 
\label{Hne}
\end{eqnarray}
where akin to earlier examples the components of the (rescaled)
effective nonlinear strain tensor $\mm{w}$ are given by
\begin{eqnarray}
  w_{zz} &=& \partial_z u_z + \frac{1}{2}
  \, (\nabla_{\perp} u_z)^2,\\
  w_{\alpha\beta} &=& \frac{1}{2} \,\left(\partial_\alpha u_\beta 
    + \partial_\beta u_\alpha
    -  \partial_\alpha u_z\partial_\beta u_z\right).
\label{wij}
\end{eqnarray}
The underlying rotational symmetry of the parent isotropic phase
guarantees a vanishing of one of the five elastic
constants\cite{GL,LMRX}, $\mu_{\perp z}$ (i.e., the $w_{z\alpha}^2$ is
missing from \rf{Hne}), that usually characterizes harmonic
deformations of a three-dimensional uniaxial solid\cite{Landau}, here
with the uniaxial axis $\zh$ chosen spontaneously.  Interestingly,
examining \rf{wij}, $w_{zz}$ and $w_{\alpha\beta}$ have structure of,
respectively, a uniaxially (e.g., smectic) and biaxially (e.g.,
columnar) periodic critical state.  Thus, nematic elastomer is a three
dimensional amorphous solid that is an amalgam of a smectic and
columnar liquid crystals.

Given this softness of the harmonic elasticity associated with
symmetry-imposed vanishing of the $\mu_{z\perp}$ modulus, thermal and
heterogeneity-driven fluctuations are divergent in
3d.\cite{XingRadzAOP,NEelSL} Concomitantly, the elastic terms
$(\grad_\perp u_z)^2$ and $(\partial_z {\bf u}_\perp)^2$ associated
with transverse distortions are missing from the quadratic part of the
elastic energy.  Thus, as in other critical matter discussed in
earlier sections, one expects qualitative importance of above elastic
nonlinearities \rf{wij} in the presence of thermal fluctuations and
network heterogeneity.\cite{XingRadzAOP,NEelSL}

Similar to their effects in smectic, columnar liquid crystals,
polymerized membranes, and other critical matter discussed earlier, in
bulk elastomers thermal fluctuations (and network heterogeneity) leads
to anomalous elasticity controlled by a tuning-free infrared
attractive fixed point, with universal length-scale dependent elastic
moduli, \cite{XingRadzAOP,NEelSL}
\begin{eqnarray}
K_{eff} &\sim& L^\eta,\ \ \mu_{eff}\sim L^{-\eta_u}.
\end{eqnarray}
The resulting critical state can be shown\cite{XingRadzAOP} to be
characterized by a universal non-Hookean stress-strain relation
\begin{equation}
  \sigma_{zz} \sim (\varepsilon_{zz})^\delta
\end{equation}
and a {\em negative} Poisson ratio for extension
$\varepsilon_{xx} > 0$ transverse to the nematic axis
\begin{eqnarray}
\varepsilon_{yy}=\frac{5}{7}\varepsilon_{xx},\ \ \
\varepsilon_{zz}=-\frac{12}{7}\varepsilon_{xx}.
\end{eqnarray}
While considerable progress has been made in understanding these
fascinating materials, many questions, particularly associated with
network heterogeneity remain open.

\section{Summary and conclusions}
In this chapter, I discussed a novel class of ordered states of matter
that I dub ``critical matter''.  The common crucial feature of these
exotic states is that, as a result of the underlying symmetry broken
in the ordered state, a subset of elastic moduli of the associated
Goldstone modes vanish identically. As a result, for sufficiently low
dimensions, such system exhibit harmonic fluctuations that are
divergent in the thermodynamic limit, and require treatment of
Goldstone modes' nonlinearities throughout the ordered phase (not just
near a critical point). Treating these within renormalization group
analysis, leads to an infrared stable fixed point that controls the
resulting highly nontrivial, strongly interacting, and universal
ordered state. The latter exhibits properties akin to that of a
critical point, but extending over the entire phase, which I therefore
naturally refer to as a ``critical phase''.

As prominent examples of such systems, I have discussed smectics (with
many soft and hard matter realizations, from paired superfluids
(FFLO), spin-orbit coupled and frustrated bosons to quantum Hall
stripes), cholesterics, helical magnets, columnar phases, putative
spontaneous vortex lattices, polymerized membranes and nematic
elastomers.  While significant progress has been made in
characterizing these systems, discovery of new systems (e.g., as
quantum examples in physically accessible dimensions), understanding
of effects of random heterogeneity and dynamics remain challenging
open problems.

I finish by noting that there are also critical phases (many appearing
as counter-parts of the ones discussed in this chapter), that I have
not discussed here, where fluctuations are driven by quenched
disorder, rather than by temperature or quantum fluctuations.  These
exhibit critical states of glassy Goldstone modes and generically
appear in dimensions higher than their thermal
counterparts\cite{RTaerogelPRL, RTaerogelPRB, BRTCaerogelScience,
  SRTcolumnar, magneticSCprl, magneticSCprb, RadzihovskyNelsonEuro,
  RadzihovskyNelsonPRA, LRflatglassPRB, MorseLubensky, LRmembraneAOP,
  LRbuckling, XingRadzAOP}, and are thus of even stronger experimental
relevance.

\section{Acknowledgments}
The material presented in these lectures is based on research done
with a number of wonderful colleagues and friends, most notably David
Nelson, John Toner, Pierre Le Doussal, Tom Lubensky, and Xiangjun
Xing. I am indebted to them for much of my insight into the material
presented here. This work was supported by the National Science
Foundation through grants DMR-1001240 and DMR-0969083 as well by the
Simons Investigator award from the Simons Foundation.

On a personal note, I have met and discussed physics with Michael
Fisher at conferences, most often at the wonderful Rutgers Stat Mech
meetings. I recall fondly Michael always sitting in a front row,
sometimes flanked by Daniel and Matthew, strikingly paying close
attention and asking probing questions even in the 3 mins student
talks, and as attentively as in invited talks by distinguished senior
faculty.  I recall him always being kind and supportive in my own
career.  Although I have never had the pleasure of working with
Michael, I have greatly benefited from his insight through his many
seminal papers that influenced my own RG work (some described in this
chapter). But perhaps more importantly, I indirectly learned physics
from Michael through David Nelson, Daniel and Matthew, all of whom
have strongly shaped my style, taste and approach to physics. And for
that I am eternally grateful.


\begin{thebibliography}{99}
\bibitem{criticalTcomment} In quantum phase transitions, temperature
  $T$ introduces a finite size cutoff $\hbar/k_BT$ into the imaginary
  time. As a result critical behavior can be probed even in a fan of
  $\delta g(T))$ away from a quantum critical
  point.\cite{SachdevQPTbook}
\bibitem{mLifshitz} A generalized $m$-Lifshitz model, has elasticity
  with $m$ ``soft'' (Laplacian) axes and $d-m$ complementary ``hard''
  (gradient) axes. In this nomenclature, the conventional classical 3d
  smectic is the $m=2$-Lifshitz model, and 2+1d quantum smectic and 3d
  classical columnar liquid crystal are described by the
  $m=1$-Lifshitz model.  Other generalizations include a nonscalar
  Goldstone-mode field as in e.g., tethered
  membranes\cite{JWSmembranes,LRmembraneAOP, RTtubulePRL, RTtubulePRE}
  and nematic elastomers\cite{XingRadzAOP}
\end{thebibliography}
\end{document}